\documentclass[12pt]{article}%

\usepackage{graphicx}%
\usepackage{multirow}%
\usepackage{amsmath,amssymb,amsfonts}%
\usepackage{amsthm}%
\usepackage{mathrsfs}%
\usepackage[title]{appendix}%
\usepackage[dvipsnames,table]{xcolor}
\usepackage{textcomp}%
\usepackage{manyfoot}%
\usepackage{booktabs}%
\usepackage{algorithm}%
\usepackage{algorithmicx}%
\usepackage{algpseudocode}%
\usepackage{listings}%

\usepackage{hyperref}
\usepackage{mathtools}
\usepackage{enumerate}
\usepackage{physics}
\usepackage{caption}
\usepackage{subcaption}
\usepackage{tikz}
\usepackage[margin=1.in]{geometry}

\def\ds{\displaystyle}

\DeclarePairedDelimiterX{\innerp}[2]{\langle}{\rangle}{#1,#2}
\newcommand{\parenthesis}[1]{\left ( #1 \right)}

\def\R{\mathbb{R}}

\def\E{\mathbb{E}}
\def\V{\mathbb{V}}

\renewcommand{\vec}[1]{{\mathchoice
                     {\mbox{\boldmath$\displaystyle{#1}$}}
                     {\mbox{\boldmath$\textstyle{#1}$}}
                     {\mbox{\boldmath$\scriptstyle{#1}$}}
                     {\mbox{\boldmath$\scriptscriptstyle{#1}$}}}}

\usepackage{cite}
\title{\LARGE \bf
Control Synthesis with Reinforcement Learning: A Modeling Perspective}

\author{Nikki Xu$^1$ \and Hien Tran$^1$}
\date{%
    $^1$Department of Mathematics, North Carolina State University\\[2ex]%
}
\begin{document}
\maketitle
\abstract{
Controllers designed with reinforcement learning can be sensitive to model mismatch. We demonstrate that designing such controllers in a virtual simulation environment with an inaccurate model is not suitable for deployment in a physical setup. Controllers designed using an accurate model is robust against disturbance and small mismatch between the physical setup and the mathematical model derived from first principles; while a poor model results in a controller that performs well in simulation but fails in physical experiments. Sensitivity analysis is used to justify these discrepancies and an empirical region of attraction estimation help us visualize their robustness. 
}

\maketitle

\section{Introduction}

Balancing a single inverted pendulum on a cart is a hallmark control task that has been studied in many settings. In textbooks, a simplified model of this system is often used to study key concepts in optimal control such as reachability, controllability, stability, etc. Engineers use this system to learn how to design a controller, and traditional control schemes such as PID, MPC, and LQR all work very well and can be implemented in a physical system \cite{miranda_application_nodate, ozana_design_2012, kennedy_real-time_2016, jezierski_comparison_2017, abeysekera_modelling_2018}. However, controller tuning can often be time-consuming and potentially cause lots of wears-and-tears on the physical systems, not to mention safety concerns in some cases. To avoid these problems, most of the literature is devoted to designing controllers in simulations or virtual environments. However, it is understood that no models are perfect replicas of the world, so controls that are completely designed from simulation should always be tested to be robust to noise, disturbance, and model mismatch. In recent years, advances in machine learning, and in particular, reinforcement learning, have been gaining tremendous success, creating new avenues towards controller designs \cite{riedmiller_neural_2005, mattner_learn_2012, lillicrap_continuous_2019}. 

The connection between control and reinforcement learning (RL) has long been recognized \cite{sutton_reinforcement_1992, lewis_optimal_2012,polydoros_survey_2017, recht_tour_2019}, but industries are often hesitant to directly adopt control schemes that are the result of the reinforcement learning algorithm, often citing safety concerns. Brunke et al.  \cite{brunke_safe_2022} provided a rather comprehensive review on reinforcement learning that addresses safe robotics deployment in the real world, which is sometimes called robust reinforcement learning. Morimoto and Doya \cite{morimoto_robust_2005} established a framework to design a controller using an RL environment that explicitly accounts for disturbance and modeling errors, but assumed the convexity of the reward function and the monotonicity of the policy functions. Nguyen et al. \cite{nguyen_robust_2021} provided a certificate for the closed-loop stability of a neural network controlled system, but the open-loop system must be linear. Berkenkamp et al. \cite{berkenkamp_safe_2017} and Richards et al. \cite{richards_lyapunov_2018} constructed candidate Lyapunov functions with neural networks to certify the stability of a system, but were more focused on designing an algorithm that expands the estimated region of attraction for safe exploration. Petsagkourakis et al. \cite{petsagkourakis_chance_2022} proposed a modified policy gradient method to guaranty, with high probability, that a set of safety constraints is not violated; however, their approach involves expensive Monte Carlo simulations and Bayesian optimizations. Anderson et al. \cite{anderson_tight_2023} provided robustness certification of ReLU networks for control purposes, but only considered disturbance of the input and not any model mismatch. 

Traditional controller design often involves a tuning phase, whereas RL-based controllers typically lack this step of physical verification, and when naively deploying virtually trained RL-based controllers, the result can be destructive. The difference between control performance in simulation and in reality is sometimes referred to as ``sim-to-real gap'' \cite{zhao_sim--real_2020}. We propose using a simple local sensitivity analysis on model parameters to help guide practitioners in reducing this gap. Local sensitivity reveals the effect of model parameters on the dynamical system without adding significant computational cost. Using a high fidelity model for training has clear advantages due to its inherent small sim-to-real gap \cite{sontakke_residual_2023, aoshima_examining_2025}, although domain randomization over a low fidelity model pair with recurrent neural network for dynamics inference has been shown to be successful with sim-to-real transfer \cite{peng_sim--real_2018}. A more comprehensive review of domain randomization in RL for robotics control can be found in \cite{muratore_robot_2022}. Ramos et al. \cite{ramos_bayessim_2019} proposed a method for domain randomization based on Bayesian statistics, which is a very insightful framework to quantify the uncertainties associated with model parameters, even though it can be computationally expensive for high dimensional systems. There is also a considerable amount of effort that goes into sim-to-real transfer in a computer vision setting \cite{du_auto-tuned_2021, chattopadhyay_augcal_2024, marougkas_integrating_2025}, which will be an interesting future direction to consider. 

In this paper, our aim is to design a controller with a neural network using reinforcement learning and investigate its robustness to a potential mismatch between the physical setup and the mathematical model derived from the first principles. These models are used as reinforcement learning environments, in place of the standard out-of-the-box cartpole environments in the Python Gymnasium package. We demonstrate that designing such controllers in a virtual simulation environment with an inaccurate model is not suitable for deployment in a physical setup. The organization of this paper is as follows: Section 2 covers the background materials needed to develop the models for simulation, reinforcement learning algorithm used for controller design, as well as the formulas for local sensitivity analysis; Section 3 contains our comparative study of the ``sim-to-real'' gap as observed in the lab; Section 4 is a local sensitivity analysis of the controlled systems along with a region of attraction estimation for the 3 types of controlled systems using a simple Monte Carlo simulation.

\section{Preliminary}
\label{section:prelim}
In this section, we discuss some background materials for modeling a single inverted pendulum on a cart in a physical laboratory setting, as well as using the REINFORCE algorithm to learn a continuous controller parameterized by a 2-layer ReLU neural network. 

\subsection{System Dynamics}
\label{section:prelim_sys_dyn}
\begin{figure}[h!]
    \begin{subfigure}{0.4\textwidth}
        \centering
        \includegraphics[scale=0.07]{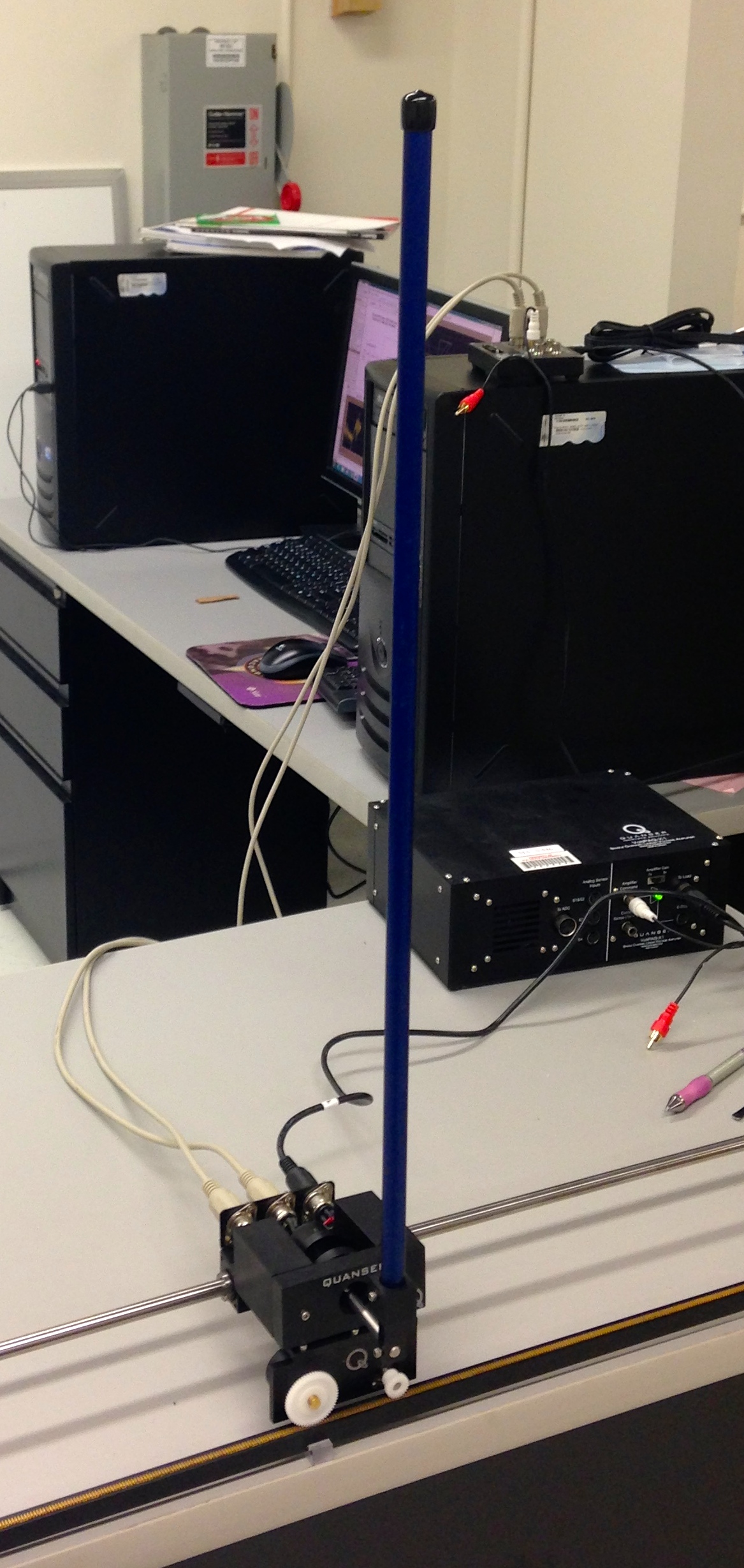}
        \caption{Picture of the Inverted Pendulum in the Lab}
        \label{fig:lab}
    \end{subfigure}%
    \begin{subfigure}{0.59\textwidth}
        \begin{tikzpicture}[scale=1.0] 
    \fill[gray!30] (-3,-0.4-0.15) rectangle (4,-0.2-0.15);
    \draw[thick] (-3,-0.2-0.15) -- (4,-0.2-0.15);

    \fill[gray!60] (-1,-0.2) rectangle (1,0.5);
    \draw[thick] (-1,-0.2) rectangle (1,0.5);
    
    \fill[black] (-0.5,-0.2) circle (0.15);
    \fill[black] (0.5,-0.2) circle (0.15);
    
    \draw[ultra thick] (0,0.5) -- (1,2.8);
    
    \fill[black] (1,2.8) circle (0.2);
    
    \node[left] at (1-0.2,2.8) {$m_p$};
    \node at (0,0.18) {$m_c$};
    
    \draw[->] (2,0.2) -- (2,0.8) node[right] {$y$};
    \draw[->] (2,0.2) -- (2.6,0.2) node[below] {$x$};
    
    \draw[->, thick] (0.51,1.65) arc (40:75:1.1);
    \node[below right] at (0,1.9) {$\alpha$};
    
    \draw[->, thick] (-1.8,0.2) -- (-1,0.2);
    \node[above] at (-1.3,0.2) {$F_c$};

\end{tikzpicture}
    \centering
    \caption{Schematic Diagram of the Pendulum on a Cart}
    \label{fig:Sketch}
    \end{subfigure}
\caption{The Inverted Pendulum in the Lab and Its Schematic Diagram}
\end{figure}

The equation of motion that corresponds to the inverted pendulum in our physical lab shown in Figure~\ref{fig:lab} was derived in \cite{kennedy_real-time_2016}. A schematic diagram of the system is shown in Figure~\ref{fig:Sketch}. The positive direction is defined as the cart moving to the right and the pendulum rotating counterclockwise. The equation of motion is derived from a Lagrangian mechanics first principle approach and is given by
\begin{subequations}
\label{eq:Lagrangian}
\begin{align}
     & \dv{ }{t} \parenthesis{\pdv{L}{\dot{x}}} - \pdv{L}{x} = F_c - B_{c} \dot{x}, \\
     & \dv{ }{t} \parenthesis{\pdv{L}{\dot{\alpha}}} - \pdv{L}{\alpha} = -B_{p} \dot{\alpha},
\end{align}
\end{subequations}
where $L$ is the Lagrangian function 
\begin{multline}    
    \label{eq:langrangian_total}
    L = \frac{1}{2} \parenthesis{m_c \dot{x}^2 
        + J_m \parenthesis{
            \frac{K_g}{r_{mp}} \dot{x}
            }^2 
        + m_p \parenthesis{
                \dot{x}^2 - 2l_p \cos(\alpha) \dot{\alpha} \dot{x} + l_p^2 \dot{\alpha}^2
                } 
        + \frac{1}{3} m_p l_p^2 \dot{\alpha}^2 }
        \\
        - m_p g l_p \cos(\alpha),
\end{multline}
$F_c$ is the driving force of the motor, $B_{c}$ is the damping coefficient of horizontal movement of the cart, and $B_p$ is the damping coefficient of rotation of the pendulum. The motor driving force is supplied by a voltage $V_m$, between $[-10,10]$ volts, and this ultimately is the control variable in our system. The cart position, $x$, is zero in the middle of the track, and the pendulum angle, $\alpha$, is zero in the upright position. The variables are listed in Table~\ref{TableOfVars}, and the parameters along with the physical meaning of each symbol used in the model are listed in Table~\ref{TableOfParam}. 
\begin{table}[h!]
    \centering
    \begin{tabular}{c|c|c}
    \hline
        Variable Description & symbol & Unit \\
    \hline
    \hline
        Position of Cart & $x$ & m\\
        Velocity of Cart & $\dot{x}$ & m/s \\
        Acceleration of Cart & $\ddot{x}$ & m/s$^2$ \\
        Angle of Pendulum & $\alpha$ & rad \\
        Angular Velocity of Pendulum & $\dot{\alpha}$ & rad/s \\
        Angular Acceleration of Pendulum & $\ddot{\alpha}$ & rad/s$^2$ \\
        Voltage Applied to  Motor & $V_m$ & V \\
    \hline
    \hline
    \end{tabular}
    \caption{List of Variables}
    \label{TableOfVars}
\end{table}

\begin{table}[h!]
    \centering
    \begin{tabular}{c|c|c}
    \hline
        Parameter Description & Symbol & Value \\
    \hline
    \hline
        Mass of Cart & $m_c$ & 0.94 kg\\
        Mass of Pendulum & $m_p$ & 0.23 kg\\
        Length of Pendulum & $l_p$ & 0.3302 m\\
        Gravitational Acceleration & $g$ & 9.8 m/s$^2$\\
        Radius of Motor Pinion & $r_{mp}$ & $6.35 \cdot 10^{-3}$ m \\
        Motor Armature Resistance  & $R_m$ & 2.6 $\Omega$ \\
        Equivalent Viscous Damping at Hinge & $B_p$ & 0.0024 N-m-s/rad \\
        Equivalent Viscous Damping on Motor Pinion & $B_c$ & 5.4 N-m-s/rad \\
        Gear Ratio & $K_g$ & 3.71 \\
        Electromotive Force (EMF) Constant & $K_m$ & $7.67 \cdot 10^{-3}$ V-s/rad \\
        Motor Torque Constant & $K_t$ & $7.67 \cdot 10^{-3}$ N-m/A \\
        Rotational Moment of Inertia of the Motor’s Output Shaft & $J_m$ & $3.90 \cdot 10^{-7}~kg-m^2$ \\
        Pendulum’s Moment of Inertia at Hinge & $J_p$ & $3.344 \cdot 10^{-2}$ ~ kg-m$^2$ \\
    \hline
    \hline
    \end{tabular}
    \caption{List of Parameters and their Values}
    \label{TableOfParam}
\end{table}
Finally, to reformulate the equations into a system of first order equations, we define the state variable $\vec{z} = (z_1, z_2, z_3, z_4)$, such that
    \[
    z_1 = x, z_2 = \dot{x}, z_3 = \alpha,  z_4 = \dot{\alpha}, 
    \]
thus, the dynamical system becomes
        \begin{align}
        \label{Sys2ndOrder}
            \dot{z_1} = & z_2  \nonumber
            \\
            \begin{split}
                \dot{z_2} = & 
                - \frac{3 r_{mp}^2 B_p \cos(z_3) {z_4}}{l_p D} 
                - \frac{4 m_p l_p r_{mp}^2 \sin(z_3) {z_4}^2}{D}
                - \frac{4 (R_m r_{mp}^2 B_{c} + K_g^2 K_t K_m) {z_2}}{R_m D}
                \\ 
                & + \frac{3 m_p r_{mp}^2 g \cos{(z_3)} \sin{(z_3)}}{D}
                + \frac{4 r_{mp} K_g K_t V_m}{R_m D} 
            \end{split} \nonumber
            \\
            \dot{z_3} = & z_4  
            \\
            \begin{split}
                \dot{z_4} = & 
                - \frac{3 \parenthesis{(m_c+m_p) r_{mp}^2 + J_m K_g^2} B_p {z_4}}{m_p l_p^2 D}
                - \frac{3 m_p r_{mp}^2 \cos(z_3)\sin(z_3) {z_4}^2}{D}  
                \\ 
                & - \frac{3 (R_m r_{mp}^2 B_{c} + K_g^2 K_t K_m) \cos(z_3) {z_2}}{R_m l_p D}
                + \frac{3\parenthesis{(m_c+m_p)r_{mp}^2 + J_m K_g^2} g \sin(z_3)}{l_p D}
                \\
                & + \frac{3 r_{mp}^2 \cos(z_3) K_g K_t V_m}{R_m l_p D}, 
            \end{split}
            \nonumber, 
        \end{align}
    where
    \[
    D = 4(m_c+m_p) r_{mp}^2 + 4 J_m K_g^2 + 3 m_p r_{mp}^2 \sin^2(z_3). 
    \]
    Henceforward, we will refer to the dynamical system in Equation (\ref{Sys2ndOrder}) as the lab model. 

\subsection{Simplified Linear Dynamics}
\label{section:prelim_sys_dyn_linear}
    In practice, models with such details as described in \ref{Sys2ndOrder} are not commonly implemented, and practitioners are tempted to use a simplified model. The simplest model is a linear time invariant (LTI) one, which can be obtained by making the following assumptions:
    \begin{itemize}
        \item Viscous damping is neglected, i.e., $B_p=0, B_c=0$.
        \item The rotational kinetic energy of both the cart and the pendulum is neglected, i.e. $J_m=0, \frac{1}{6}m_p l_p^2\dot{\alpha}^2=0$. 
        \item The system can be well-approximated by linearization around $\vec{0}$. 
    \end{itemize}
    Again, based on the Lagrangian mechanics principle in (\ref{eq:Lagrangian}), a simplified model based on the first two assumptions is given by
    \begin{equation}
    \label{eq:SimplePhysics}
        \begin{bmatrix}
            \ddot{x} \\ \ddot{\alpha}
        \end{bmatrix}
        = \begin{bmatrix}
            m_c + m_p & -m_p l_p \cos(\alpha) \\ -\cos(\alpha) & l_p
        \end{bmatrix} ^{-1}
        \begin{bmatrix}
            F_c - m_p l_p \sin(\alpha) \dot{\alpha}^2 \\ g \sin(\alpha)
        \end{bmatrix}. 
    \end{equation}
    The simplified dynamics can again be rewritten as a system of first order ordinary differential equations using the same state variables. Then linearizing it near $\vec{z} = (0,0,0,0)$, the simplified dynamics becomes
    \begin{align}
    \label{eq:LTI_with_Control}
        \dot{\vec{z}} & = 
        \begin{bmatrix}
            0 & 1 & 0 & 0 \\
            0 & -\frac{K_g^2 K_t K_m}{R_m r_{mp}^2 m_c} & \frac{m_p g}{m_c} & 0 \\
            0 & 0 & 0 & 1 \\
            0 & -\frac{K_g^2 K_t K_m}{R_m r_{mp}^2 m_c l_p} & \frac{m_c + m_p}{m_c l_p}g & 0
        \end{bmatrix} \vec{z}
        + 
            \begin{bmatrix}
            0 \\
            \frac{K_g K_t}{R_m r_{mp} m_c}
            \\
            0 \\
            \frac{K_g K_t}{R_m r_{mp} m_c l_p}
            \end{bmatrix}
        V_m. 
    \end{align}
Henceforward, we will refer to the dynamical system in Equation (\ref{eq:LTI_with_Control}) as the linear model. 

\subsection{Reinforcement Learning}
\label{section:prelim_RL}
In this reinforcement learning (RL) framework, the controller action $V_m$ is sampled randomly from a Gaussian distribution that is parameterized by a fully connected neural network, whose input is the observed state variable, and it outputs the mean and standard deviation of the Gaussian distribution. The control input then interacts with the discretized dynamical system to produce an observation of the next time step, and a reward is evaluated based on the new observation. More specifically, the reward is 1 at step $n$ if the observed states at step $n+1$ satisfy  
\[
z_1 \in [-0.2, 0.2] \text{ and } z_3 \in [-0.2, 0.2], 
\]
and the reward is 0 if the states are outside of these bounds. That is, the controlled cart position should be within 0.2 meters from the center of the rail, and the controlled pendulum angle should be within 0.2 radians of being upright; otherwise, the controller failed. We assume that the velocities have no constraints. The initial conditions of the dynamical systems are sampled uniformly from (-0.08, 0.08) for all four state variables, which mimics the experimental lab setup of a stationary upright pendulum on a cart that is near the center of the rail. 

We now formalize our problem in standard RL notation, adapted from Sutton and Barto \cite{sutton_reinforcement_2020}. The RL problem is formulated as a Markov decision process that is characterized by the following:
\begin{itemize}
    \item a set of states, $S$;
    \item a set of actions, $A$;
    \item a transition probability function, $P(s, a, s')$, denoting the probability that action $a$ in state $s$ will lead to state $s'$ in the next time step; 
    \item a reward function, $R(s, a, s')$, denoting the reward received after transitioning from state $s$ to $s'$ with action $a$. 
\end{itemize}
A policy $\pi : S \to A$ maps a state to a distribution of actions, and we denote $\pi(a | s)$ the probability of selecting action $a \in A$ given state $s \in S$. The sequence of actions and states, $(s_0, a_0, s_1, a_1, \dots, s_t, a_t, s_{t+1}, \dots)$, is called a trajectory, which we denote as $\tau$, and the subscripts on state and action denote the time step on the trajectory. We write $\tau \sim \pi$ to describe a trajectory $\tau$ where the action at every step is sampled according to $\pi$, that is, $a_t \sim \pi(\cdot | s_t)$; and the next state is determined by the state transition function $P: S \times A \times S \to [0,1]$, that is, the probability of obtaining the state $s_{t+1}$ given $s_t$ and $a_t$ should satisfy $\ds \text{Pr}(s_{t+1} \in S'| s_t=s, a_t=a) = \int_{S'} P(s,a,s')ds'$ for every measurable $S'\subset S$. We denote $R(\tau)$ the cumulative discounted reward of a trajectory
        \(
            R(\tau) = \sum_t \gamma^t R(s_{t+1} | s_t,a_t)
        \)
where $\gamma \in (0,1)$ is a fixed real number, and $R(s_{t+1} | s_t,a_t) $ denotes the reward for arriving at $s_{t+1}$ from $s_t$ when taking action $a_t$. 

    Suppose that the initial state is sampled from some distribution, i.e., $s_0 \sim \rho_0$. Suppose that the following states are modeled by a distribution that is entirely determined by the immediate previous state and action only, and suppose that the action is sampled from a policy that only depends on the current state, i.e. $a_t \sim \pi(\cdot | s_t)$. Then the probability of a T-step trajectory is as follows.
    \begin{equation}
    \label{eq:prob_traj}
    \Tilde{P}(\tau|\pi) = \rho_0(s_0) \prod_{t=0}^{T-1} P(s_{t+1}|s_t, a_t) \pi(a_t | s_t).     
    \end{equation}
    Then the expected reward is
    \begin{equation}
        \label{eq:exp_reward}
        J(\pi) = \int_\tau \Tilde{P}(\tau | \pi) R(\tau) = \underset{\tau \sim \pi}{\E}[R(\tau)] = 
        \underset{\tau \sim \pi}{\E} \left[ \sum_{t=0}^\infty \gamma^t R_t(s_t, a_t, s_{t+1}) \right]. 
    \end{equation}
    
We use a simple policy gradient (PG) algorithm REINFORCE \cite{towers_gymnasium_2023} to obtain a policy that maximizes the above reward. A pseudocode is provided below: 

\begin{algorithm}
\caption{REINFORCE for Continuous Control}
\label{alg:PG}
\begin{algorithmic}
\State Initialize Reward $\gets 0$
\State Initialize state randomly from uniform distribution, $s_0 \sim \text{Uniform}[-0.08, 0.08]^4$
\While{Reward $<$ Threshold}
    \State Sample $a_t \sim \pi_\theta(\cdot | s_t)$
    \State Clip $a_t \in [-10, 10]$
    \State Integrate the right hand side of Equation~\ref{Sys2ndOrder} or \ref{eq:LTI_with_Control} with semi-Euler's method to estimate $s_{t+1}$
  \If{$\abs{x} > 0.2$ or $\abs{\alpha} > 0.2$}
    \State Reward $\gets 0$ and Reset initial condition
  \Else 
    \State Reward $\gets$ Reward $+1$
  \EndIf
\EndWhile
\Repeat{ After each episode is done:}
  \State Update the loss function $\sum \ln{J} = -\sum\ln{(P(\tau|\pi))}\cdot$Reward (i.e. negative log likelihood of the expected reward in Equation~\ref{eq:exp_reward})
  \State Update parameters with one step of NAdam to maximize the expected reward
\Until{Maximum number of episodes is reached. }
\end{algorithmic}
\end{algorithm}

\section{Reinforcement Learning using Lab vs. Linear Model} 

Training a controller using reinforcement in a virtual environment before implementing the result in a real system has gained tremendous interest from practitioners in recent years, partly due to the ease of using a virtual environment such as Gymnasium \cite{towers_gymnasium_2023} and partly due to the high cost that comes with running the same trial and error experiment on a real physical system tens of thousands of times can be prohibitive \cite{bates_benchmarking_2022}. Through this study, we demonstrate that virtual training should be performed in an environment that closely mimics the physical system rather than a subpar approximation of one. 

\subsection{Training Neural Network for Control}
A simple vanilla policy gradient algorithm was used to design a controller to stabilize a single inverted pendulum on a cart. The algorithm is mainly adapted from the REINFORCE tutorial in \cite{towers_gymnasium_2023}. However, the environment was modified to sample the action from an interval rather than two discrete states because the controller action in the lab is a voltage range. Moreover, the model in the environment was also modified to describe the two dynamical systems with which we have: (i) a detailed lab model as shown in (\ref{Sys2ndOrder}), and (ii) an LTI model as shown in (\ref{eq:LTI_with_Control}). The architecture of the neural network that parameterizes the control policy distribution, $\pi_\theta$, is the same for both experiments. The optimization objective and the training algorithm are the same for both experiments: maximizing the reward using PyTorch NAdam. The hyperparameters of the PyTorch optimizer, such as the learning rate, were tuned separately. A pseudocode is described in Algorithm~\ref{alg:PG}. More specific to this project, $\pi_\theta$ is a two-hidden-layer fully connected neural network shown in Figure~2. The input is the state variables $s_t = (x, \dot{x}, \alpha, \dot{\alpha})$. This feeds into a 36-neuron hidden layer and then another 10-neuron hidden layer, both with ReLU activation functions. The widths of the layers were chosen arbitrarily. The output layer consists of two neurons representing the mean and standard deviation of a normal distribution $\mathcal{N}(\mu, \sigma)$ from which the policy action $a_t$ is sampled at each time step $t$. 

To reiterate, as seen in Algorithm~\ref{alg:PG}, while the cumulative reward is less than some threshold, at step $t$, the current state $s_t$ serves as the input layer vector, goes through the neural network and outputs two parameters that are required to specify the normal probability distribution of the actions to be taken. The policy samples an action $a_t$ from this distribution and restricts it to $[-10,10]$ volts due to the physical limitations of the motor, and uses either the lab model or the LTI model to integrate one step forward in time using the semi-Euler method with a fixed step size of $h=0.01$, thus generating the new state $s_{t+1}$. If the pendulum  remains balanced on the track, that is, the pendulum's angle is between $\pm 0.2$ radians and the cart's position is between $\pm 0.2$ meters, then $r_{t+1} = 1$ is appended to the reward list, and training continues. If $s_{t+1}$ does not meet these criteria, the training is stopped, the rewards are discounted and updated, and the weights and biases of the neural network are updated through one step of backpropagation using the NAdam optimizer in PyTorch. Finally, training is considered complete when the agent successfully balances the pendulum for 20 seconds 95\% of the time, that is, the threshold is set to $(0.95)(20/0.01) = 1950$, and the agent acts 1950 time steps without the pendulum falling or the cart going off track. 

\begin{figure}
\centering
\begin{tikzpicture}[scale=0.6, transform shape]
\label{fig:nn}
    \foreach \i in {1,...,4}
        \node[circle, draw, minimum size=.5cm] (I-\i) at (0, -\i*1.45) {$z_{\i}$};
    
    \node[circle, draw, minimum size=.5cm] (H1-1) at (3, -1) {$l_{1,1}$};
    \node at (3, -2.25) {$\vdots$};
    \node[circle, draw, minimum size=.5cm] (H1-2) at (3, -3.75) {$l_{1,j}$};
    \node at (3, -5) {$\vdots$};
    \node[circle, draw, minimum size=.5cm] (H1-3) at (3, -6.5) {$l_{1,36}$};
    
    \node[circle, draw, minimum size=.5cm] (H2-1) at (6, -1) {$l_{2,1}$};
    \node at (6, -2.25) {$\vdots$};
    \node[circle, draw, minimum size=.5cm] (H2-2) at (6, -3.75) {$l_{2,j}$};
    \node at (6, -5) {$\vdots$};
    \node[circle, draw, minimum size=.5cm] (H2-3) at (6, -6.5) {$l_{2,10}$};
    
    \node[circle, draw, minimum size=.5cm] (O-1) at (9, -2.5) {$\mu$};
    \node[circle, draw, minimum size=.5cm] (O-2) at (9, -4.5) {$\sigma$};
    
    \foreach \i in {1,...,4}
        \foreach \j in {1,...,3}
            \draw[->] (I-\i) -- (H1-\j);
    
    \foreach \i in {1,...,3}
        \foreach \j in {1,...,3}
            \draw[->] (H1-\i) -- (H2-\j);
    
    \foreach \i in {1,...,3}
        \foreach \j in {1,...,2}
            \draw[->] (H2-\i) -- (O-\j);
    
    \node[align=center] at (0,1) {Input\\Layer};
    \node[align=center] at (3,0) {Hidden\\Layer 1};
    \node[align=center] at (6,0) {Hidden\\Layer 2};
    \node[align=center] at (9,1) {Output\\Layer};
\end{tikzpicture}
    \caption{Architecture of the Neural Network Parameterizing a Normal Distribution for Policy. Inputs are state variables, outputs are $\mu$ and $\sigma$, denoting the mean and variance, respectively, of the normal distribution from which the actions are sampled from at each time step.}
\end{figure}

\subsection{Training Results and Simulations}
\begin{figure}[h!]
     \centering
     \begin{subfigure}[b]{0.49\textwidth}
         \centering
         \includegraphics[width=\textwidth]{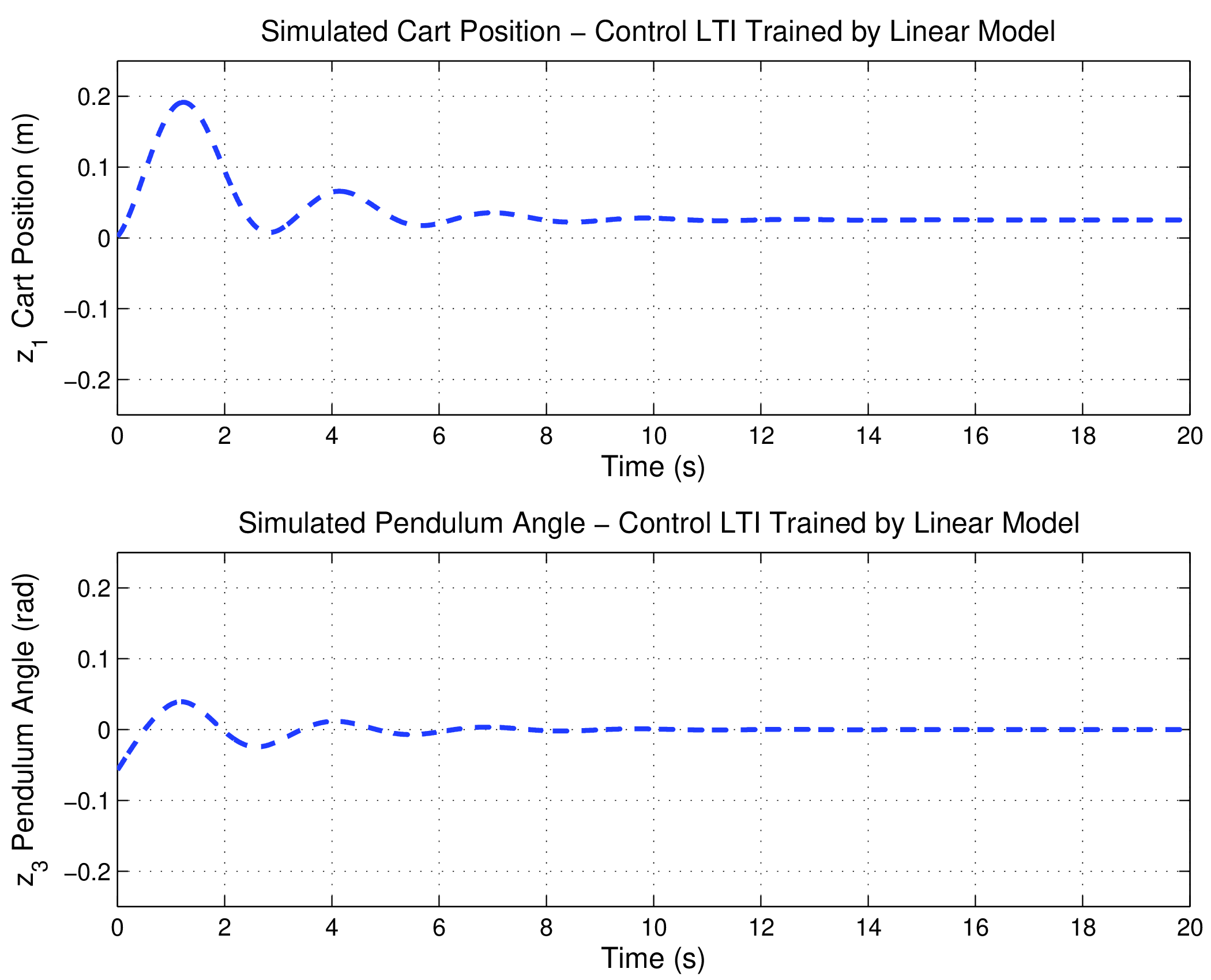}
         \caption{Trained Controller with the LTI Model Control on the Linear Model Simulation}
         \label{fig:LinControl_OnLTISys}
     \end{subfigure}
     \hfill
     \begin{subfigure}[b]{0.49\textwidth}
         \centering
         \includegraphics[width=\textwidth]{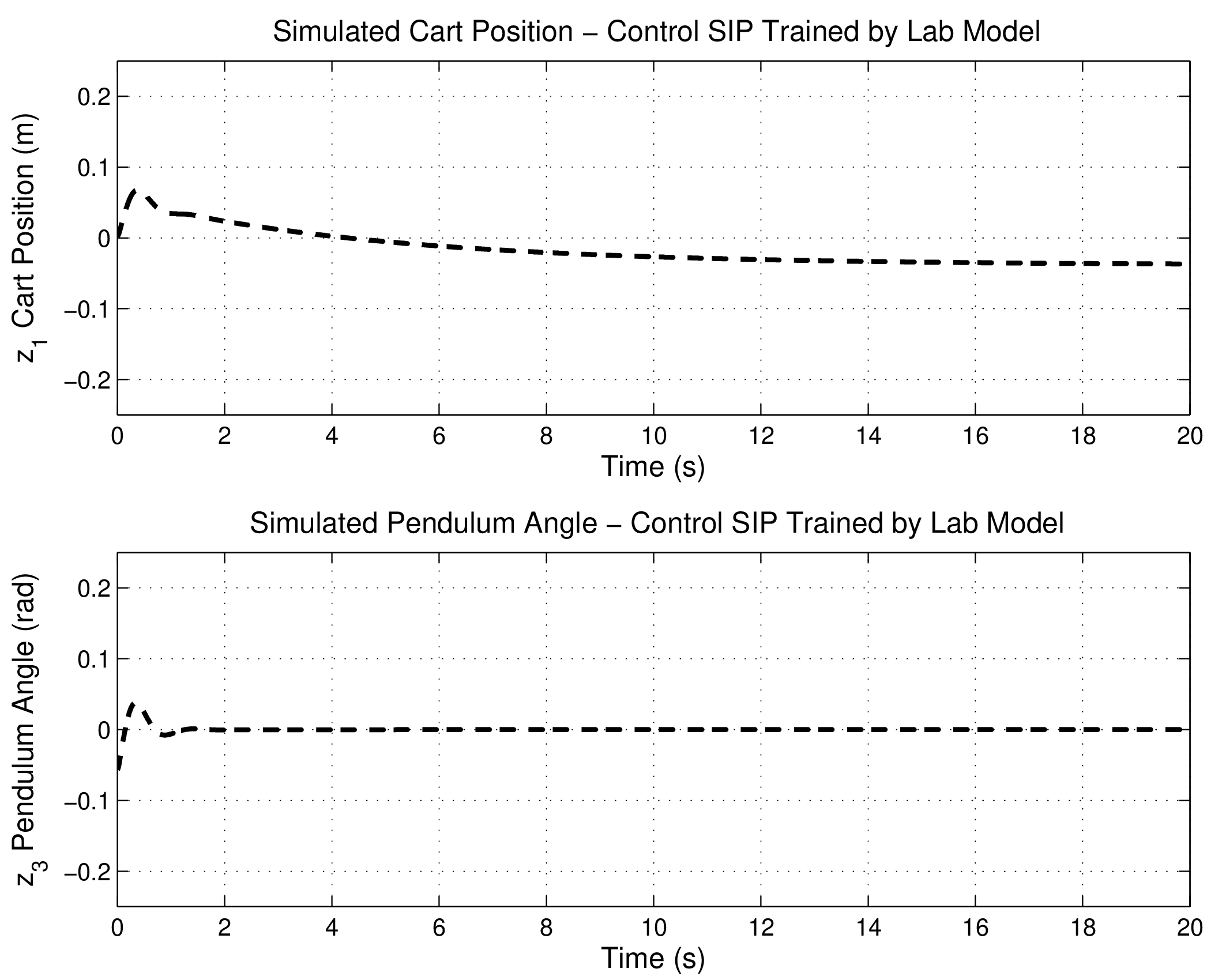}
         \caption{Trained Controller with the Lab Model Control on the Lab Model Simulation}
         \label{fig:NonLinControl_OnFullSys}
     \end{subfigure}
        \caption{Control Learned with Different Systems}
    \label{fig:Control_OnDiffSys}
\end{figure}

\begin{figure}[h!]
     \centering
     \begin{subfigure}[b]{0.49\textwidth}
         \centering
         \includegraphics[width=\textwidth]{plots/Simulations/Simulation_LTI_withLinCtrl.png}
         \caption{Trained Controller with the LTI Model Control on LTI Simulation}
         \label{fig:LinControl_OnLTISim}
     \end{subfigure}
     \hfill
     \begin{subfigure}[b]{0.49\textwidth}
         \centering
         \includegraphics[width=\textwidth]{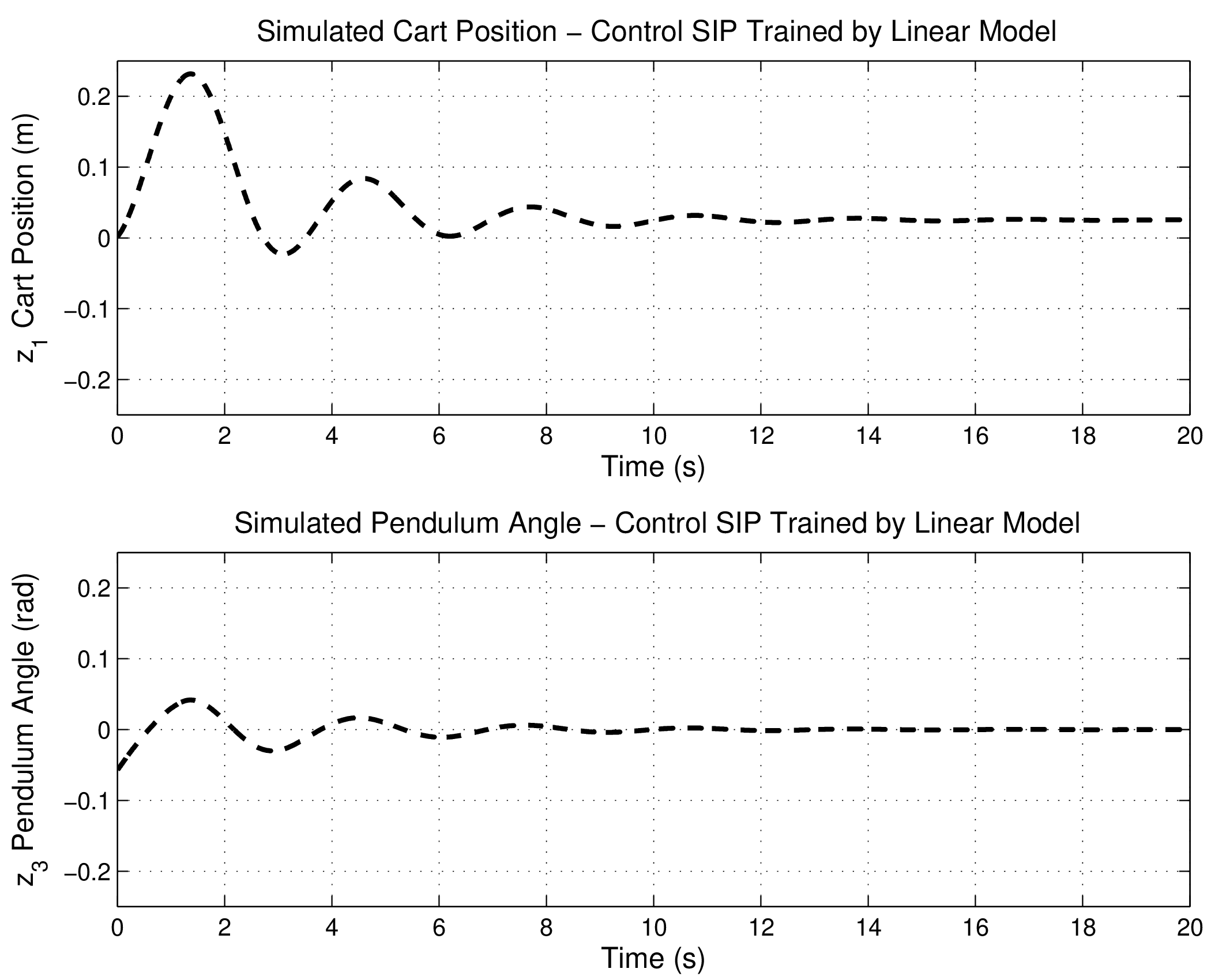}
         \caption{Trained Controller with the LTI Model Control on the Lab Model Simulation}
         \label{fig:LinControl_OnSIPSim}
     \end{subfigure}
        \caption{Control Learned with LTI Systems Simulated with Difference Systems}
    \label{fig:LinControl_OnDiffSys}
\end{figure}

Training with both LTI and lab models successfully produced controllers that can stabilize the system in simulation. In Figure~\ref{fig:Control_OnDiffSys}, we see that the controller learned using the lab model is able to stabilize the pendulum very quickly compared to the controller learned using the LTI model. Moreover, simulation also shows that the ReLU controller learned from the LTI model should be able to stabilize the pendulum when it is implemented in both the LTI model itself and the lab model, as shown in Figure~\ref{fig:LinControl_OnDiffSys}. The difference in the trajectories of the controlled systems does not appear to be significant. That is, once a controller is designed, simulating its control performance on the LTI model or the lab model yields more or less the same trajectory, demonstrating that the linearized model (Equation~\ref{eq:LTI_with_Control}) should theoretically be a good approximation of the physics model (Equation~\ref{Sys2ndOrder}). However, notice that applying the controller trained with the LTI model on the lab model yields a slightly higher oscillation of cart position, pushing it past the safety region of operations, above the 0.2 m constraint. This subtle difference can be seen between Figure~\ref{fig:LinControl_OnLTISys} and in Figure~\ref{fig:LinControl_OnSIPSim}. 


The lab implementation of this controller is deterministic, that is, we use the weights and biases of all the neurons except for the final output of standard deviation, so the controller is just a piecewise affine function of the input. The fully trained neural network controller is as follows 
    \begin{align}
    \label{eq:control_implemented}
        V_m/10 = b_m + W_m
            \parenthesis{
            \sigma \parenthesis{
                b_2 + W_2
                \sigma \parenthesis{
                    b_1 + W_1 {z}
                    }
                }
            }, 
    \end{align}
    where
    $\sigma : \R \to \R : x \mapsto \max\{0, x \}$ is a scalar function applied pointwise to a vector; $b_m$ is the bias of the mean output of the final layer; $W_m$ are the weights of the mean output of the final layer; $b_i$ and $W_i$ are the biases and weights of the hidden layer $i$, respectively; and $\vec{z}$ is the state variable $(x, \dot{x}, \alpha, \dot{\alpha})$; and $V_m$ is the voltage applied to the cart's motor. More explicitly, 
    \[
    V_m = 10 \parenthesis{ b_m + W_m \cdot \max(\vec{0}, b_2 + W_2 \cdot (\max(\vec{0}, b_1 + W_1 \cdot \vec{z}))) }. 
    \]

\subsection{Lab Experiments}

We recorded five experiments for both of the trained controllers implemented as feedback control laws on the physical pendulum system in the lab. In each experiment, the cart is stationary near the center of the track, and the pendulum is slowly raised from the resting position to the upright position. We include the plots of all the experimental trajectories in the Appendix. 

\subsubsection{Controller Trained with LTI Model}
In 3 of the 5 experiments, the controller was unsuccessful in keeping the pendulum upright, and in the other two experiments, the pendulum and the cart oscillated quite rapidly and never seemed to reach a stable position. These results are recorded in Figure~\ref{fig:Lab_LTI}. 

In the experimental run \# 1 the pendulum falls in 5 seconds and the cart simply was unable to control the initial condition, as shown in \ref{fig:LinControl_run1}. In run \#2, the cart went off track in about 4 seconds and the pendulum was never quite stable, as shown in \ref{fig:LinControl_run2}. 

In runs \#3 and \#4, the pendulum was held upright for more than 40 seconds; however, we observe that the controller actions were quite rapid and the cart oscillated a lot even without any external disturbances, which can be seen in Figure~\ref{fig:LinControl_inLab_experiments3&4}. Furthermore, the cart was very shaky in both experiments, the motor was constantly going back and forth doing too much adjustment, and we can see that the pendulum is also shaking from the same plots. 

In run \#5, the cart was able to balance the pendulum for almost 30 seconds, but the pendulum was never quite stable. As shown in Figure~\ref{fig:LinControl_inLab_experiments5}, during the first 15 seconds, the pendulum appeared to be in good balance, but then the cart began to go farther and farther away from the center and eventually falls off the track. 

In none of these five experiments did we attempt to disturb the balancing in any way, and we see that the controller trained with an inaccurate model may be very sensitive to noise. 

\subsubsection{Controller Learned from the Lab Model}

In comparison, the controller learned with the more detailed lab model not only stabilizes the physical pendulum successfully but is also robust against disturbance. As shown in Figure~\ref{fig:NonLinearModel_Experiment}, the controller was able to quickly stabilize the pendulum when it was gently tapped up to approximately a 14 degree disturbance. In particular, the spikes observed in these plots correspond to the pendulum being tapped. We observe that the controller performs better against external disturbances that are in the counterclockwise direction, but poorer from the other direction. In particular, we observed that the cart is slow in restoring the pendulum to the upright position when tapped clockwise, but much faster in coming back from the other direction. In Figure~\ref{fig:FullModel_Experiment}, we see these effects more clearly on each individual experiment. 

\section{Sensitivity Analysis}
While it isn't surprising that training in an overly simplified simulation environment does not result in a transferrable control policy, we attempt to analyze this discrepancy with a sensitivity analysis on the model parameters of the close-loop systems. More explicitly, we implement the neural network controllers trained with the LTI model and with the lab model as feedback controllers for the dynamical system described in Equation~\ref{Sys2ndOrder}. With fixed initial conditions, we can simulate the trajectories of both closed-loop systems, and then perform a derivative-based global sensitivity analysis. Under some boundedness assumptions, derivative-based sensitivity measures have been shown to be used to bound global variance-based sensitivity indices such as the Sobol indices \cite{sobol_derivative_2009, kucherenko_derivative-based_2016}. 

Recall that for a given dynamical system with arbitrary initial conditions,
\begin{equation}
\label{eq:general_IVP}
    \dv{\vec{z}}{t} = \vec{f}(t, \vec{z}(t), \vec{p}), \quad \vec{z}(t_0) = \vec{z}_0, 
\end{equation}
we can define a sensitivity matrix $\vec{s}_{jk}$ for the $j$-th state variable $\vec{z}_j$ with respect to the $k$-th parameter $\vec{p}_k$ near the nominal parameter values $\vec{p}^*$ as
\begin{equation}
\label{eq:sensitivity_matrix}
    \vec{s}_{jk}(\vec{p}^*) = \pdv{\vec{z}_j}{\vec{p}_k} \cdot \vec{p}^* . 
\end{equation}

For an analytic function $\vec{z}$, we can approximate this partial derivatives with the following complex-step method:
\begin{equation}
    \label{eq:complex_diff}
    \vec{s}_{jk} = 
    \pdv{\vec{z}_j}{\vec{p}_k}~(\vec{p}^*) \approx 
    \frac{\textbf{Im}(\vec{z}_j(\vec{p}^* + ih \vec{e}_k ))}{h},
\end{equation}
where $i = \sqrt{-1}$, and $\textbf{Im}$ denotes the imaginary part of the complex-valued function. Note that by the analytic assumption, the step size can be reduced to machine precision and usually has negligible effect on the accuracy of the approximation \cite{banks_complex-step_2015}. 

\subsection{Local Sensitivity of Model Parameters}

We first examined the setting of the uncontrolled case $V_m = 0$, in order to identify the sensitive parameters during the modeling phase; then, we simulated the same systems with trained controllers and used the same method to examine the sensitivities of the controlled systems to model parameters. 

For the uncontrolled systems, we first compare our complex-step approximation of each sensitivity matrix with the Jacobians computed using the reverse-mode auto-differentiation in JAX \cite{jax2018github}. For three randomly chosen initial conditions, we found that the relative mean square errors of the simulated sensitivity trajectories over 1 second are less than $10^{-5}$. For nonlinear systems, the RMSE of each sensitivity matrix trajectories are less than $10^{-7}$, and the errors are slightly higher for the linear system due to its divergence nature as we neglected all frictions in the system. We remark that the complex-step size $h$ does not affect the accuracy of the approximation, hence the numerical limitation largely depends on the ODE solver's accuracy. 

Since the damping coefficients, $B_p$ and $B_c$ were both neglected, the states of the uncontrolled LTI system naturally diverges when the pendulum is allowed to free-fall from near the upright position. Clearly, by omitting the damping parameters, the system dynamics is fundamentally changed from a physically realistic one, in which the pendulum eventually settles down to the downward equilibrium point, to a forever rotating system where the cart also moves infinitely far away. Moreover, due to the diverging states, we examine only the sensitivities of model parameters during the first second of simulation. While the simplification assumptions appear to be valid because the neglected parameters exhibit relatively low sensitivities, the underlying dynamical systems are fundamentally changed. When compared to the lab model, the local sensitivity trajectories of the LTI model become much more heavily influenced by the masses of the cart the pendulum as well as by gravity. Global sensitivity measures therefore become a necessity to analyze the systems since gradients alone fail to capture the overall impact of these parameters on the dynamics. 

\begin{figure}[htbp]
    \centering
    \includegraphics[width=0.99\linewidth]{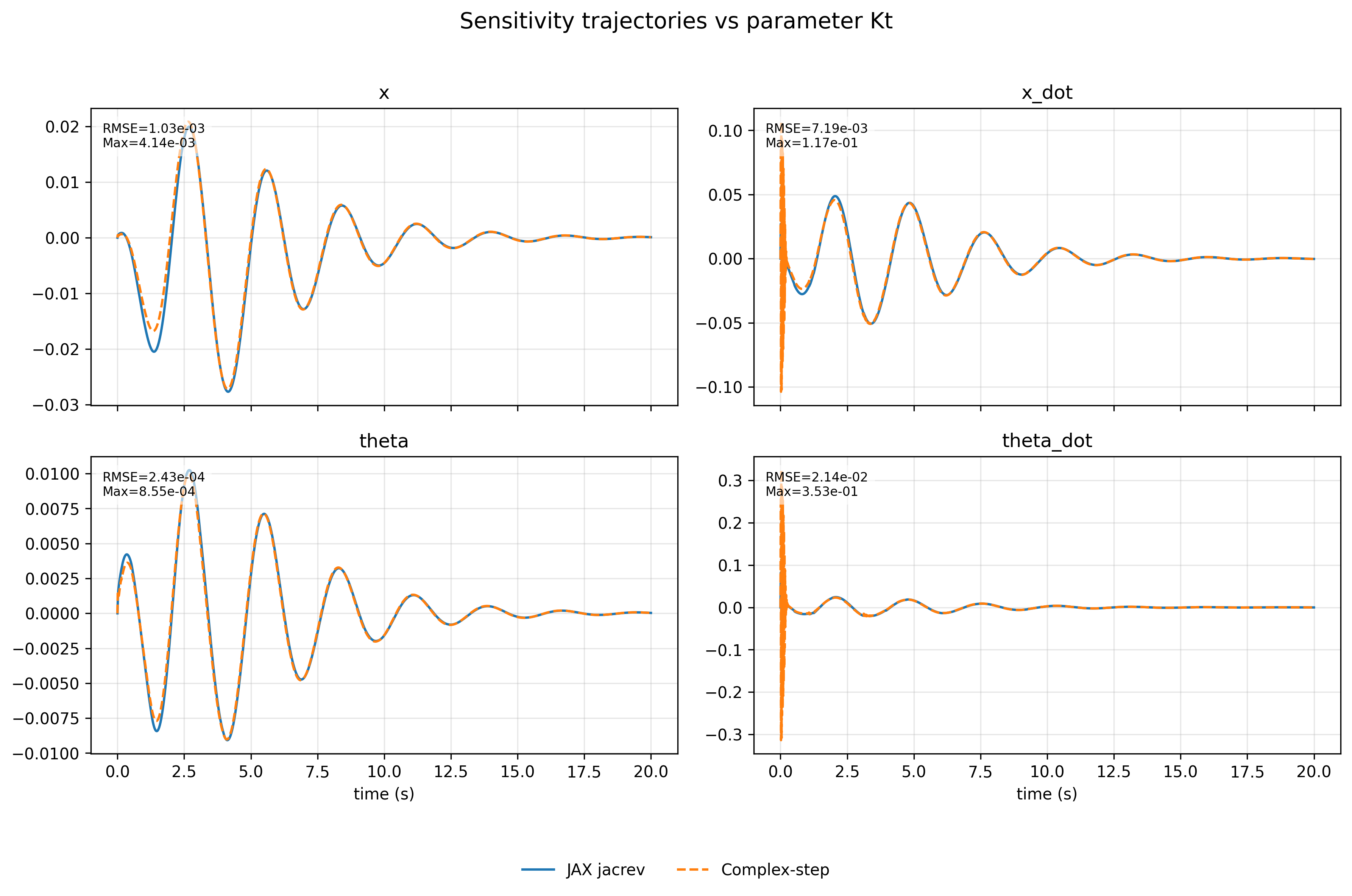}
    \caption{Sensitivity Trajectory of Model Parameter $K_t$ in the Controlled Linear System}
    \label{fig:sensitivity_controlled_lin_lin_Kt}
\end{figure}

Before we analyze the global sensitivities, we also computed the local sensitivity trajectories using both the complex-step method and reverse-mode auto-differentiation with JAX for three controlled systems:
        \begin{enumerate}
            \item {Control the linear model using the controller learned with the LTI model}
            \item {Control the lab model using the controller learned with the LTI model}
            \item {Control the lab model using the controller learned with the lab model}
        \end{enumerate}
We notice a significant improvement in computational speed using our complex-step method in all the simulation scenarios that we experimented with, reducing computation time as much as 5 times when compared to JAX. However, since our neural controllers with ReLU activation functions introduce non-smoothness to the dynamical system, and consequently non-smoothness to the sensitivity matrices as well, the complex-step method is not suitable for their approximations, resulting in much higher relative errors. In fact, if we recall the LTI model in (\ref{eq:LTI_with_Control}), non-smoothness can be observed directly from the control-affine system:
    \begin{align*}
            \dot{\vec{z}} & = 
            \begin{bmatrix}
                0 & 1 & 0 & 0 \\
                0 & -\frac{K_g^2 K_t K_m}{R_m r_{mp}^2 m_c} & \frac{m_p g}{m_c} & 0 \\
                0 & 0 & 0 & 1 \\
                0 & -\frac{K_g^2 K_t K_m}{R_m r_{mp}^2 m_c l_p} & \frac{m_c + m_p}{m_c l_p}g & 0
            \end{bmatrix} \vec{z}
            + 
                \begin{bmatrix}
                0 \\
                \color{red}{
                \frac{K_g K_t}{R_m r_{mp} m_c}
                }
                \\
                0 \\
                \color{red}{
                \frac{K_g K_t}{R_m r_{mp} m_c l_p}
                }
                \end{bmatrix}
            V_m,
    \end{align*}
where $V_m$ is the controller determined by the neural network using ReLU as the activation function as described by Equation~\ref{eq:control_implemented}. This directly contradicts the assumption for the complex-step method which requires the function to be holomorphic, therefore, one should be cautious about its applications. A clear example of this phenomenon can be observed in Figure~\ref{fig:sensitivity_controlled_lin_lin_Kt}, where the sensitivity of the controlled linear system's cart velocity and pendulum velocity with respect to the model parameter $K_t$ computed using the complex method displayed in orange dashed lines deviate from the JAX-based solutions rather significantly at the beginning of the trajectory. 

\subsection{Connection with Global Sensitivity}
Although a gradient-based local sensitivity analysis reveals how the system responds to small perturbations of model parameters, there has also been some work to demonstrate its connections to variance-based global sensitivity analysis, which aims at quantifying these responses statistically \cite{sobol_derivative_2009,kucherenko_derivative-based_2016,alexanderian_variance-based_2020}. Specifically, for a scalar function $f$ whose output depends on some uncertain parameters, $\vec{p}$, as well as on time, $t$, i.e.,
\[
Y = f(t, \vec{p}), \quad t \in [0,T], 
\]
where $\vec{p} = (p_1, \dots, p_m) \in \R^m$ is a vector of uncertain model parameters, quantifying the impact of $\vec{p}$ on $Y$ is highly dependent on efficient sampling and often incurs prohibitively high computational cost. The total effect index, denoted by $S_k^{tot}$, gives the total variance in $Y$ caused by the $k$-th parameter, ${p}_k$, as well as its interactions with any of the other parameters. This index is computed as 
\[
S_k^{tot} = 1 - \frac{\V[\E[Y|\vec{p}_{\sim k}]]}{\V[Y]}, 
\]
where $\vec{p}_{\sim k}=(p_1, \dots, p_{k-1}, p_{k+1}, \dots, p_m)$ denotes all the parameters except $p_k$; $\V[\cdot]$ and $\E[\cdot]$ denote the variance and expectation operators, respectively. 

A gradient-based method such as the one presented in the previous subsection can be used to bound the variance-based global sensitivity measures under some mild assumptions\cite{sobol_derivative_2009, kucherenko_derivative-based_2016}. More specifically, suppose that each model parameter $\vec{p}_k$ is uniformly distributed on the unit hypercube, and suppose that $f$ is a finite-time integral of the solution of a state variable $\vec{z}_j$ of our dynamical system for a fixed initial condition, and further assume that $\pdv{\vec{z}_j}{\vec{p}_k}$ is square-integrable for each $j,k$, then it is possible to use a derivative-based global sensitivity measure (DGSM) as a proxy for the total Sobol index of the parameter $k$ for variable $j$. 

\begin{figure}
    \centering
    \includegraphics[width=0.9\linewidth]{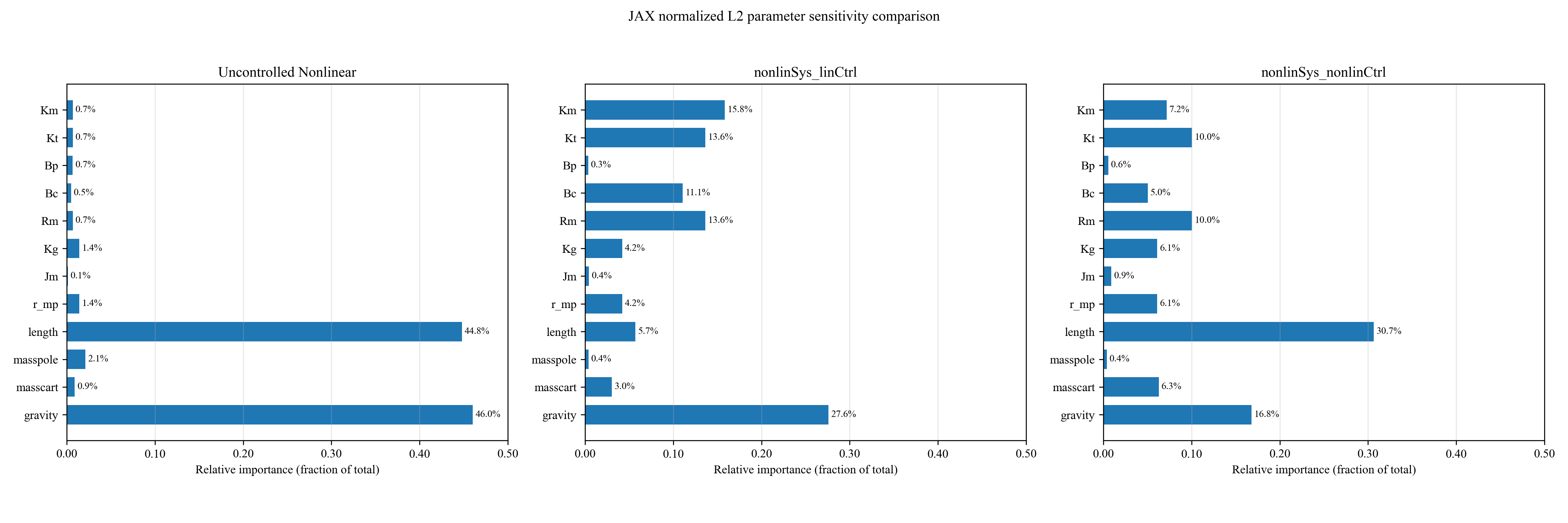}
    \caption{Comparison of the sensitivities of each parameter with various controllers}
    \label{fig:sensitivity-l2-cost}
\end{figure}

There is a vast amount of literature on global sensitivity analysis and uncertainty quantification and it is not the focus of our current study, but it is worth mentioning since the complex-method gradient calculation is fast and accurate for smooth functions; incorporating it into the time-dependent framework can be potentially helpful for robust designs of RL-based controllers in the future. As a preliminary demonstration, we approximate a derivative-based global sensitivity measure (DGSM) for the closed-loop systems, where the feedback controllers are the training results from the previous section. As shown in Figure~\ref{fig:sensitivity-l2-cost}, when the system is uncontrolled, the trajectories of the state variables are heavily influenced by the pendulum's length as well as gravity, which is consistent with physical intuition. However, different control policies are able to reduce the impact of certain parameters, and being able to quantify this impact may be beneficial for shaping the behavior of RL-based control systems. 

\begin{table}[htpb]
    \centering
    \begin{tabular}{c|c|c}
    \hline
        Parameter & LTI-based Controller & Lab-based Controller \\
    \hline
    \hline
        \rowcolor{blue!30} $m_c$ & 9 & 6 \\
        $m_p$ & 10/11 & 12 \\
        \rowcolor{blue!30} $l_p$ & 6 & 1 \\
        $g$ & 1 & 2 \\
        $r_{mp}$ & 7/8 & 7/8 \\
        $R_m$ & 3/4 & 3/4 \\
        \rowcolor{red!30} $B_p$ & 12 & 11 \\
        \rowcolor{red!50} $B_c$ & 5 & 9 \\
        $K_g$ & 7/8 & 7/8 \\
        \rowcolor{blue!30} $K_m$ & 2 & 5 \\
        $K_t$ & 3/4 & 3/4 \\
        \rowcolor{red!30} $J_m$ & 10/11 & 10 \\
    \hline
    \hline
    \end{tabular}
    \caption{Model Parameter Sensitivity Ranking in Different Control Systems}
    \label{tab:sensitivity_rankings}
\end{table}

The closed-loop control systems displayed different sensitivity rankings for some of the model parameters. The ranking of the parameter sensitivity measures from Figure~\ref{fig:sensitivity-l2-cost} are summarized in Table~\ref{tab:sensitivity_rankings}. 
For example, among the three parameters omitted during model reduction, the equivalent damping coefficient on the cart, $B_c$, ranks 9th in the closed-loop system with the controller trained with the lab model; compared to ranking \#5 for system with the controller trained with the LTI model. The other two omitted parameters, $B_p$ and $J_m$ both remain insignificant with or without control of either kind, suggesting that neglecting them may have been a decent choice. The damping parameters absent from the LTI system, $B_c$ and $B_p$, are estimated from experiments numerically, so in reality, they might deviate from the estimated values. Our sensitivity analysis suggests that the parameter, $B_c$, may be one of the main reasons why the LTI-trained controller failed in the lab experiments. Moreover, the EMF constant parameter, $K_m$, ranks \#2 in the closed-loop simulation with the controller trained with the LTI model, whereas it ranks \#5 in the system with the controlled trained with the lab model. Although this particular motor-related constant is provided by the equipment manufacturer, there are still uncertainties associated with it, so that a small deviation from its nominal value can still significantly impact the trajectory of the closed-loop system with the LTI-trained controller in lab experiments. 

\subsection{Region of Attraction}
\label{section:sensi_ICs}
Finally, we analyze the region of attraction of each of the following controlled systems:
\begin{enumerate}[(a)]
    \item Control the LTI system using the control policy learned with the LTI model
    \item Control the lab model using the control policy learned with the LTI model
    \item Control the lab model using the control policy learned with the lab model
\end{enumerate}

The idea is to see if the last policy (c) would be more ``successful'' than learning with a simplified model. So, we uniformly sample the initial conditions in the rectangle $D = [-0.2, 0.2] \times [-10, 10] \times [-0.2, 0.2] \times [-10,10] \subset \R^4$. With those initial conditions, we simulated the 3 controlled systems each for 40 s. The system is considered to be successfully controlled if the pendulum angle is within 0.05 radians from upright in the final second and if at no time during the simulation, the cart position is more than 0.2 m away from the center of the track. The latter criterion can be viewed as a safety constraint in the RL-based control system, which is an active research field \cite{berkenkamp_safe_2017, richards_lyapunov_2018, liang_accelerated_2018, petsagkourakis_chance_2022, tambon_how_2022}. 

We initially only sampled 5000 initial conditions in a rather large region, so the ROA estimation is not statistically significant. Therefore, we use the previously validated initial conditions as the center, and progressively increase the radii of their neighborhoods in order to pursue an upper bound of the basin of attraction. As a result, we end up with 80,979 total samples of initial conditions that meet the aforementioned criteria. Of the 80,979 samples, 
\begin{enumerate}[(a)]
    \item Control the LTI system using the control policy learned with the LTI model: 26,583/80,979 are stable (roughly 33\%); 
    \item Control the lab system using the control policy learned with the LTI model: 16,318/80,979 are stable (roughly 20\%); 
    \item Control the lab system using the control policy learned with the lab model: 46,178/80,979 are stable (roughly 57\%).
\end{enumerate}

\begin{figure}[htbp]
    \centering
    \includegraphics[width=0.9\linewidth]{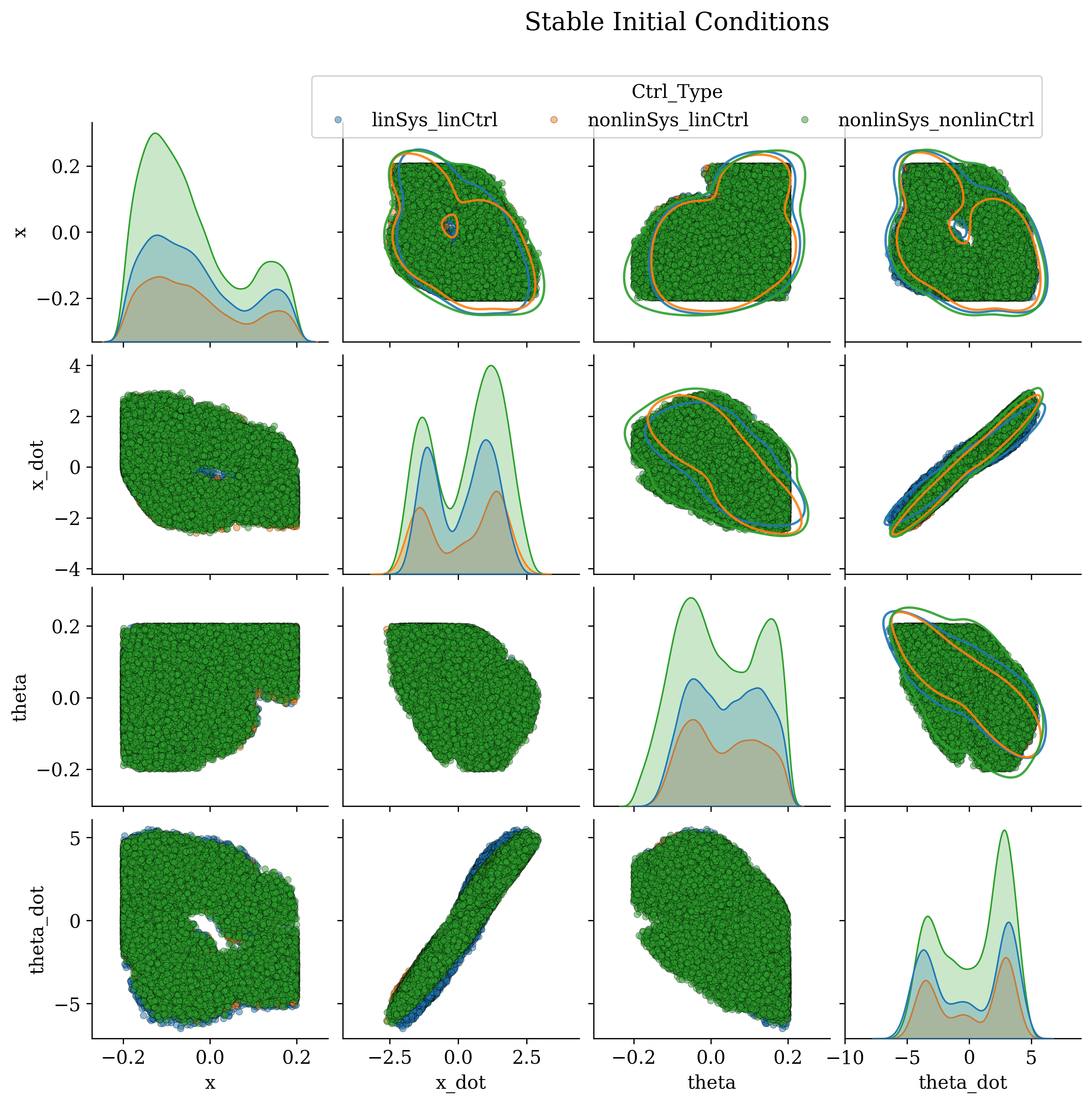}
    \caption{Region of Attraction: Sampled in Neighborhoods Around Previously Found Valid Initial Conditions}
    \label{fig:ROA_more_samples}
\end{figure}

\begin{figure}[htpb]
     \centering
     \begin{subfigure}[b]{0.49\textwidth}
         \centering
         \includegraphics[width=\textwidth]{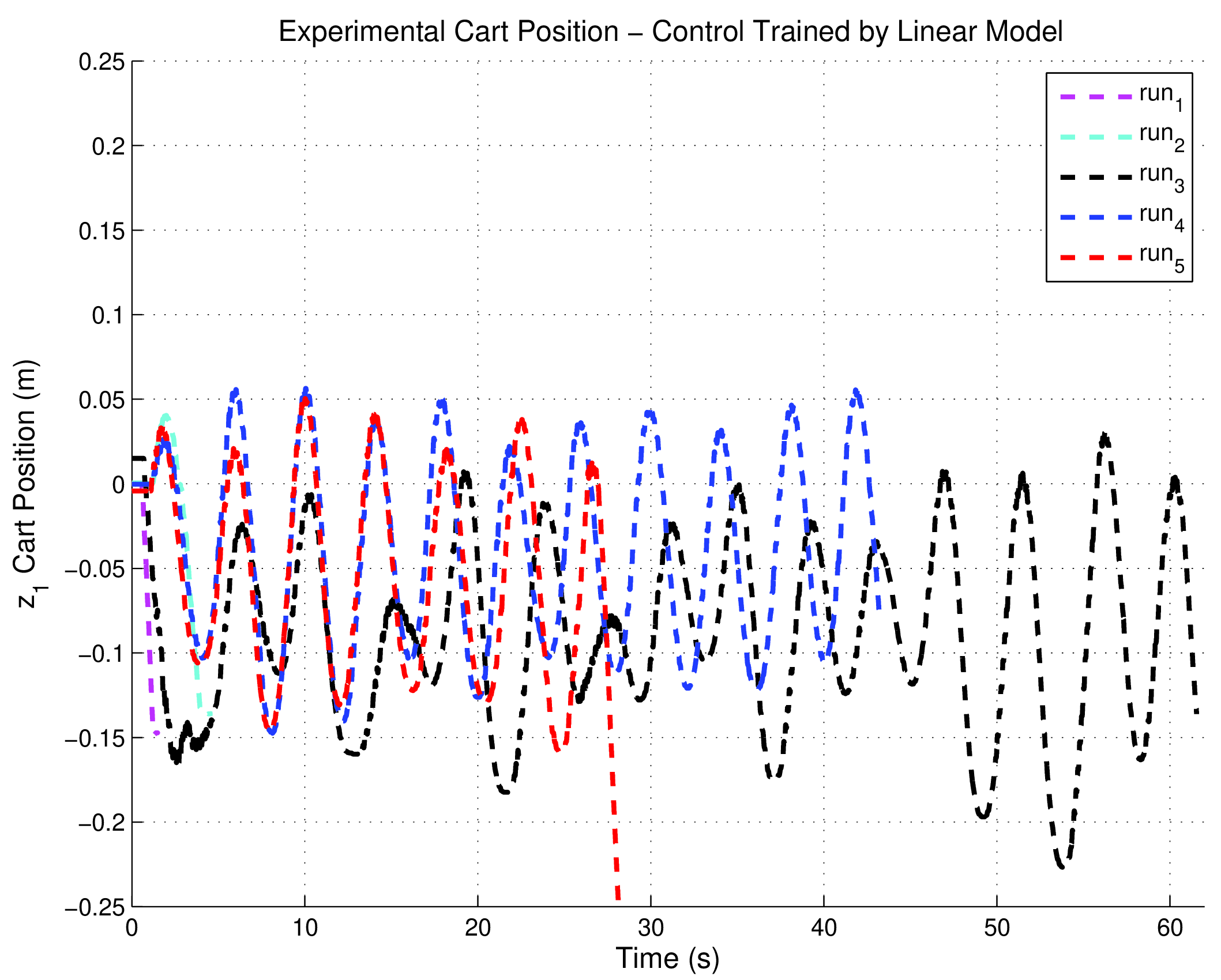}
         \caption{Cart Position}
         \label{fig:LTI_in_Lab_x}
     \end{subfigure}
     \hfill
     \begin{subfigure}[b]{0.49\textwidth}
         \centering
         \includegraphics[width=\textwidth]{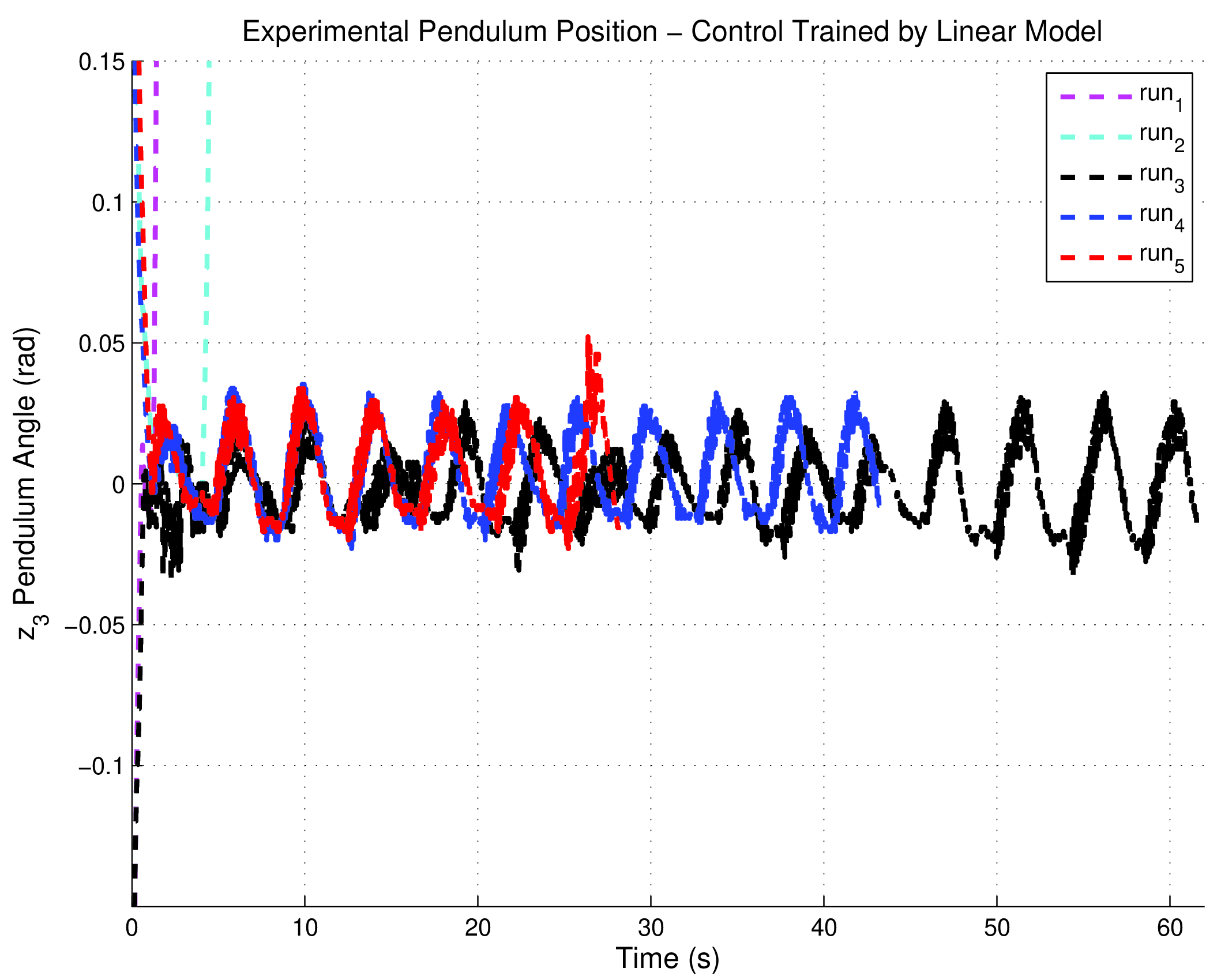}
         \caption{Pendulum Position}
         \label{fig:LTI_in_Lab_alpha}
     \end{subfigure}
        \caption{Controller Learned with LTI Model Generally Fails in Lab Experiment}
        \label{fig:Lab_LTI}
\end{figure}

\begin{figure}[htpb]
     \centering
     \begin{subfigure}[b]{0.49\textwidth}
         \centering
         \includegraphics[width=\textwidth]{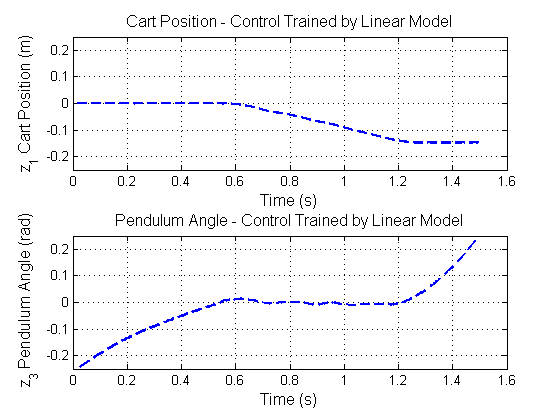}
         \caption{run \#1: Fails}
         \label{fig:LinControl_run1}
     \end{subfigure}
     \hfill
     \begin{subfigure}[b]{0.49\textwidth}
         \centering
         \includegraphics[width=\textwidth]{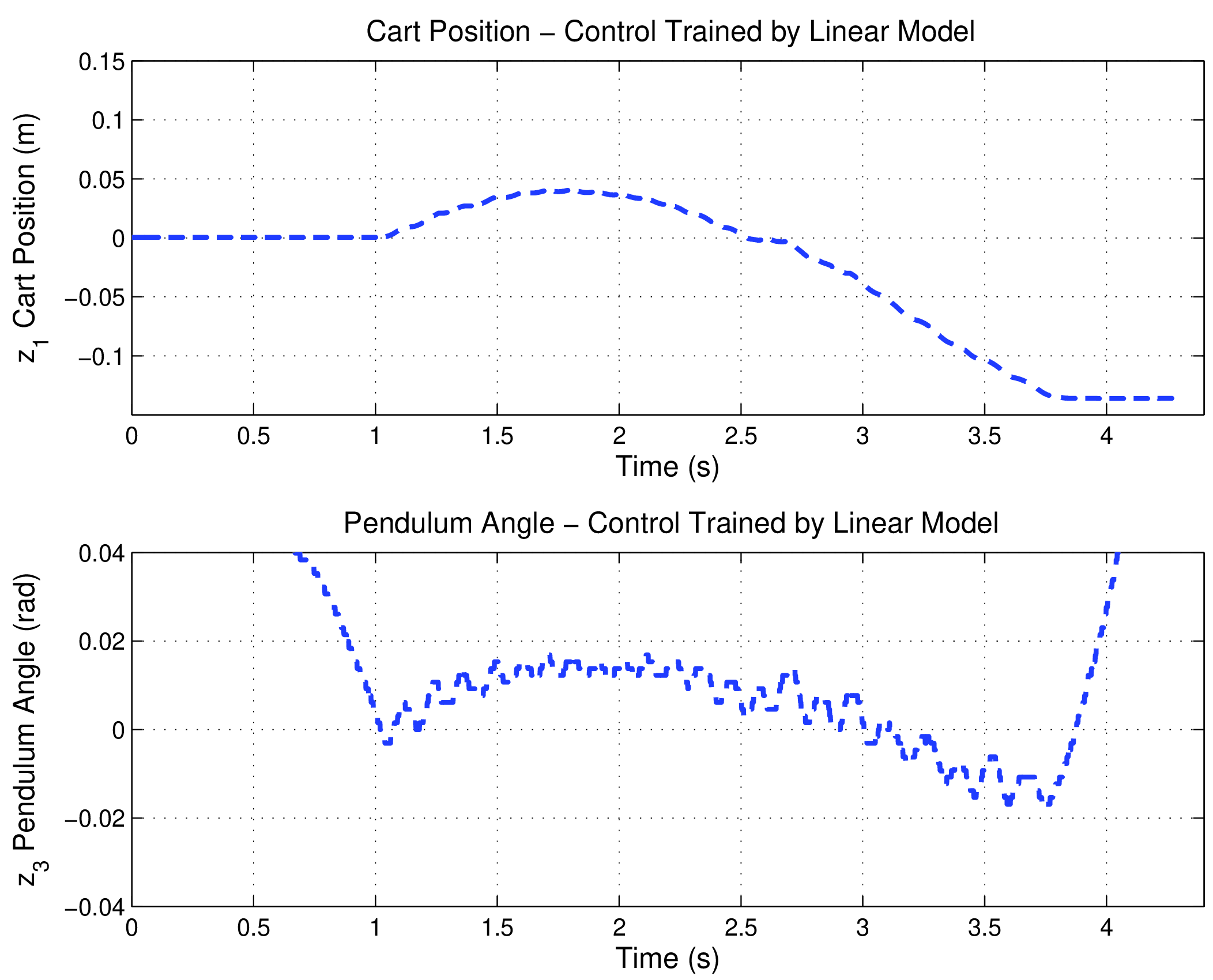}
         \caption{run \#2: Fails}
         \label{fig:LinControl_run2}
     \end{subfigure}
        \caption{Control Learned with LTI Systems Tested in the Lab Quickly Fails}
    \label{fig:LinControl_inLab_experiments1&2}
\end{figure}

\begin{figure}[htbp]
     \centering
     \begin{subfigure}[b]{0.49\textwidth}
         \centering
         \includegraphics[width=\textwidth]{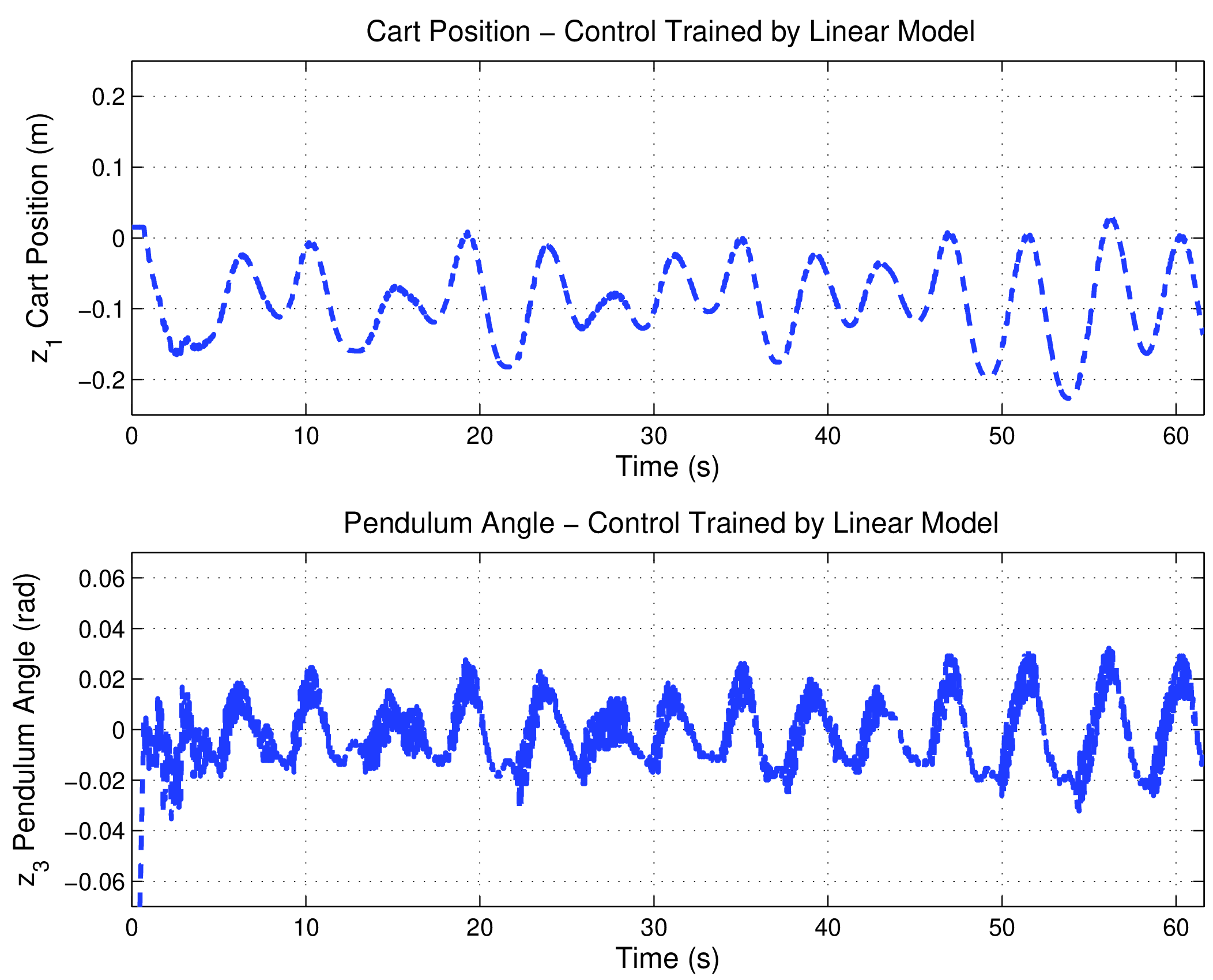}
         \caption{run \#3: Balances the Pendulum}
         \label{fig:LinControl_run3}
     \end{subfigure}
     \hfill
     \begin{subfigure}[b]{0.49\textwidth}
         \centering
         \includegraphics[width=\textwidth]{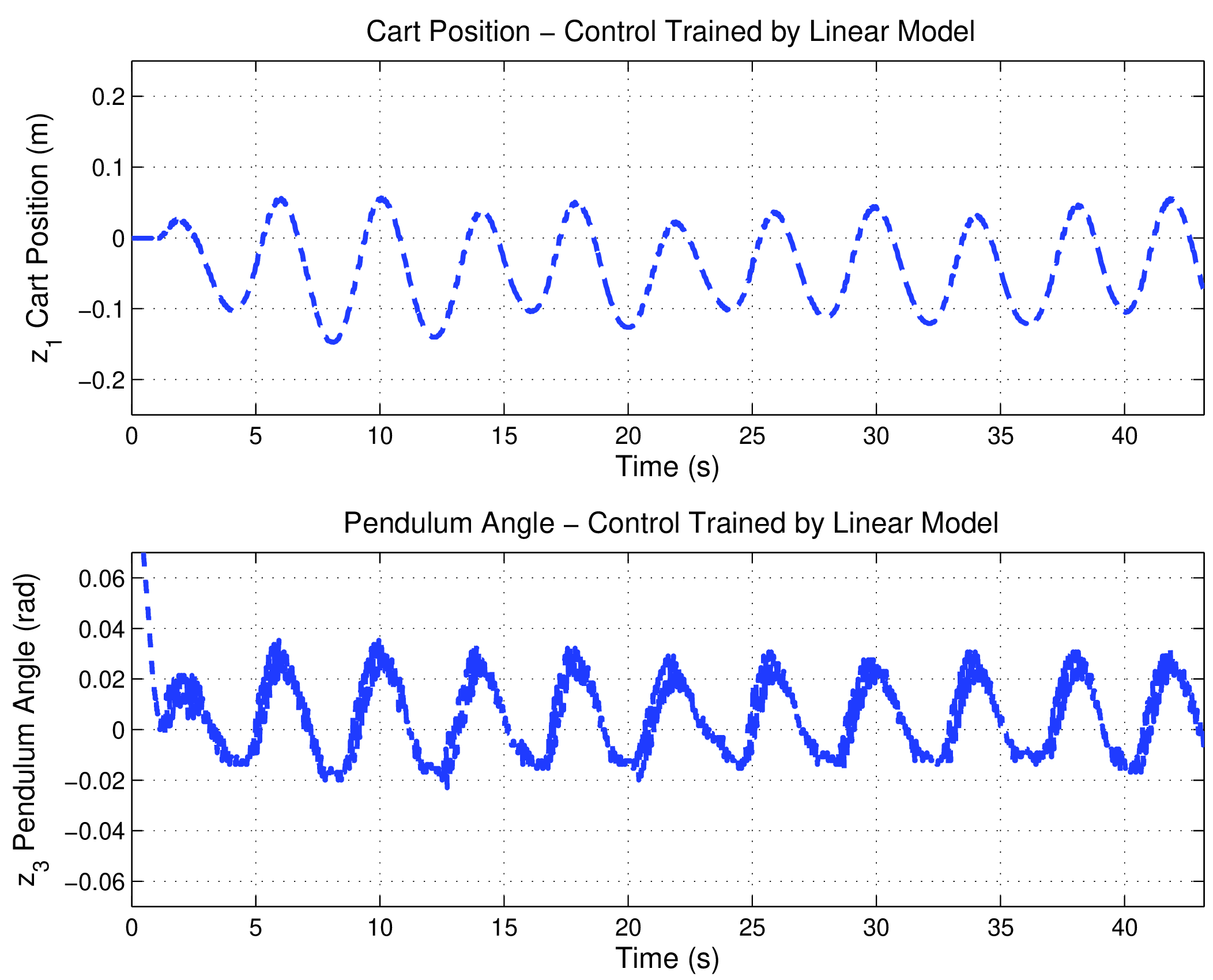}
         \caption{run \#4: Balances the Pendulum}
         \label{fig:LinControl_run4}
     \end{subfigure}
        \caption{Control Learned with LTI Systems Tested in the Lab Balances the Pendulum for over 40 sec but with a Lot of Oscillations on the Cart}
    \label{fig:LinControl_inLab_experiments3&4}
\end{figure}

\begin{figure}[htpb]
    \centering
    \includegraphics[width=0.5\linewidth]{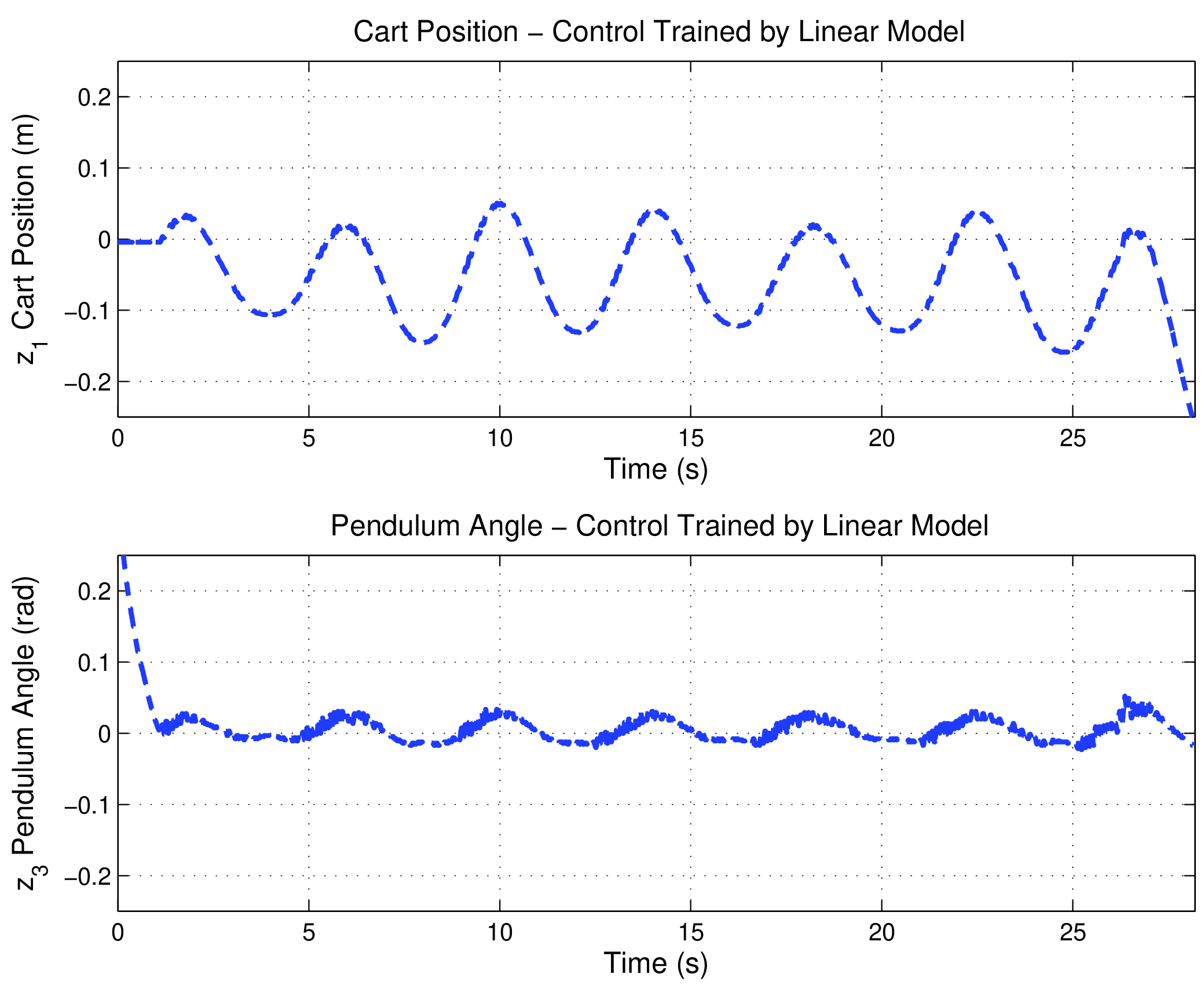}
    \caption{run \#5: Oscillation Amplified and Eventually Fails to Balance the Pendulum}
    \label{fig:LinControl_inLab_experiments5}
\end{figure}

\begin{figure}[htbp]
     \centering
     \begin{subfigure}[b]{0.49\textwidth}
         \centering
         \includegraphics[width=\textwidth]{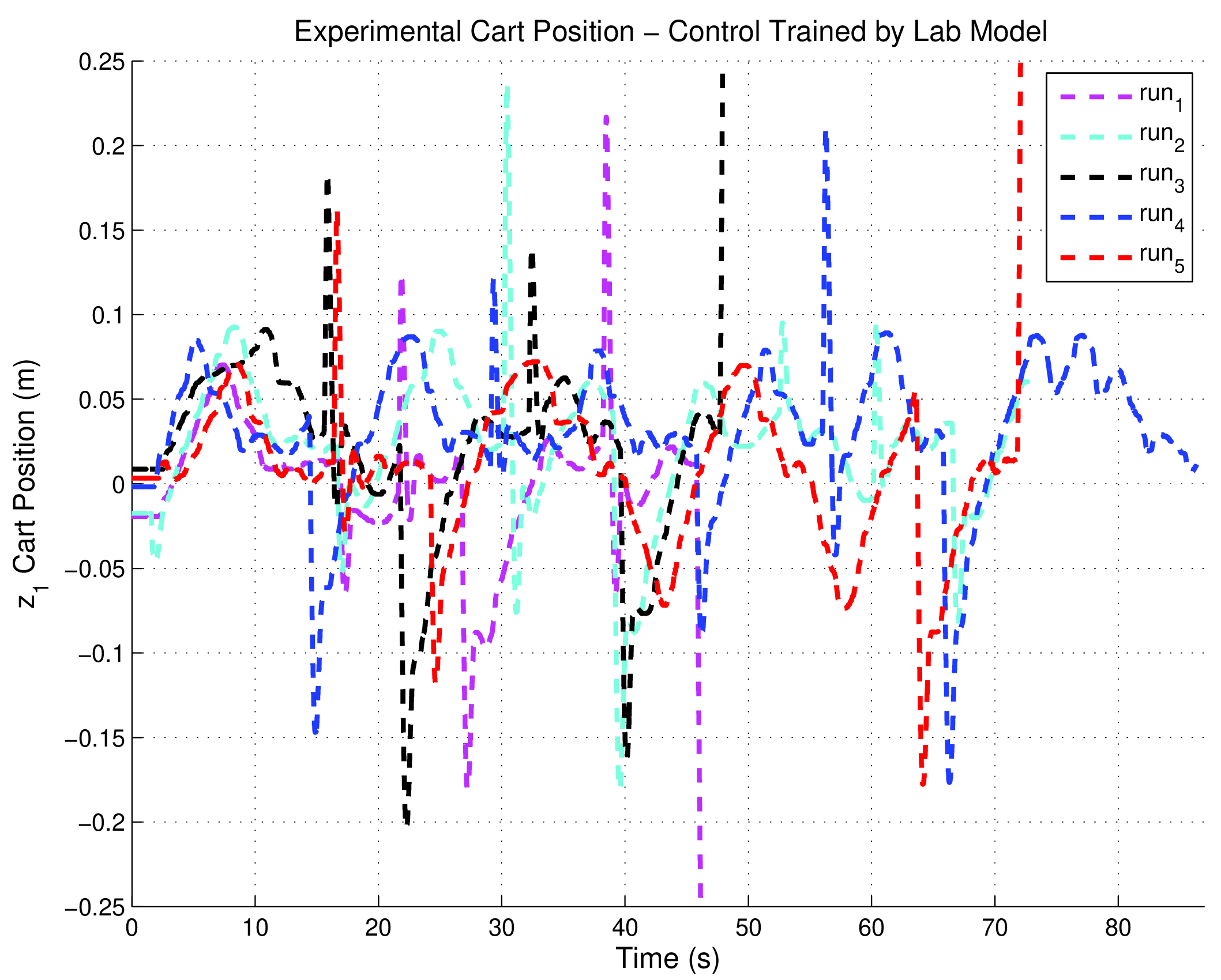}
         \caption{Cart Position}
         \label{fig:NonLinearModel_x_Experiment}
     \end{subfigure}
     \begin{subfigure}[b]{0.49\textwidth}
         \centering
         \includegraphics[width=\textwidth]{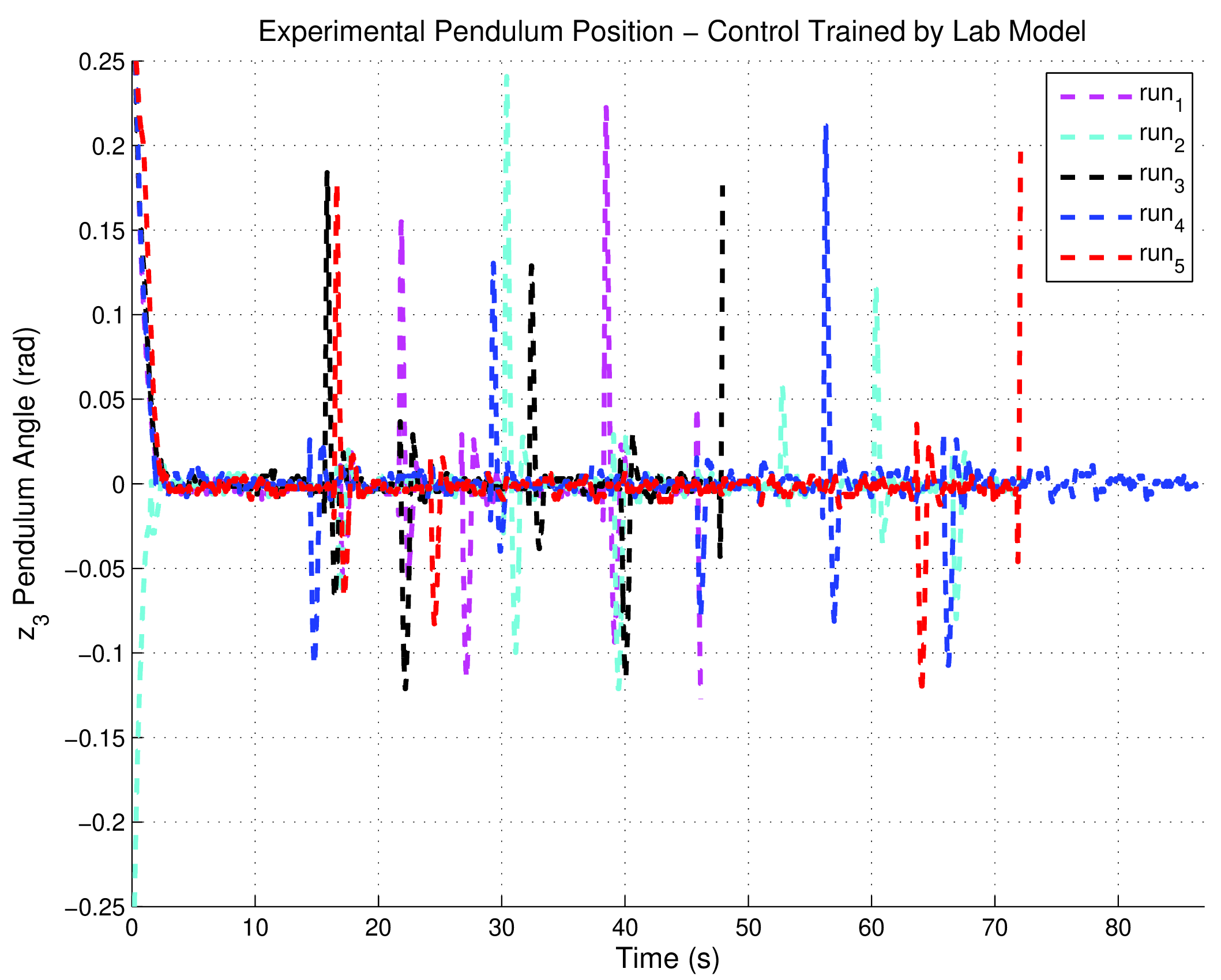}
         \caption{Pendulum Position}
         \label{fig:NonLinearModel_alpha_Experiment}
     \end{subfigure}
        \caption{Control Learned with the Lab Model Tested on the Physical Pendulum in the Lab}
        \label{fig:NonLinearModel_Experiment}
\end{figure}

\begin{figure}[htbp]
     \centering
     \begin{subfigure}[b]{0.49\textwidth}
         \centering
         \includegraphics[width=\textwidth]{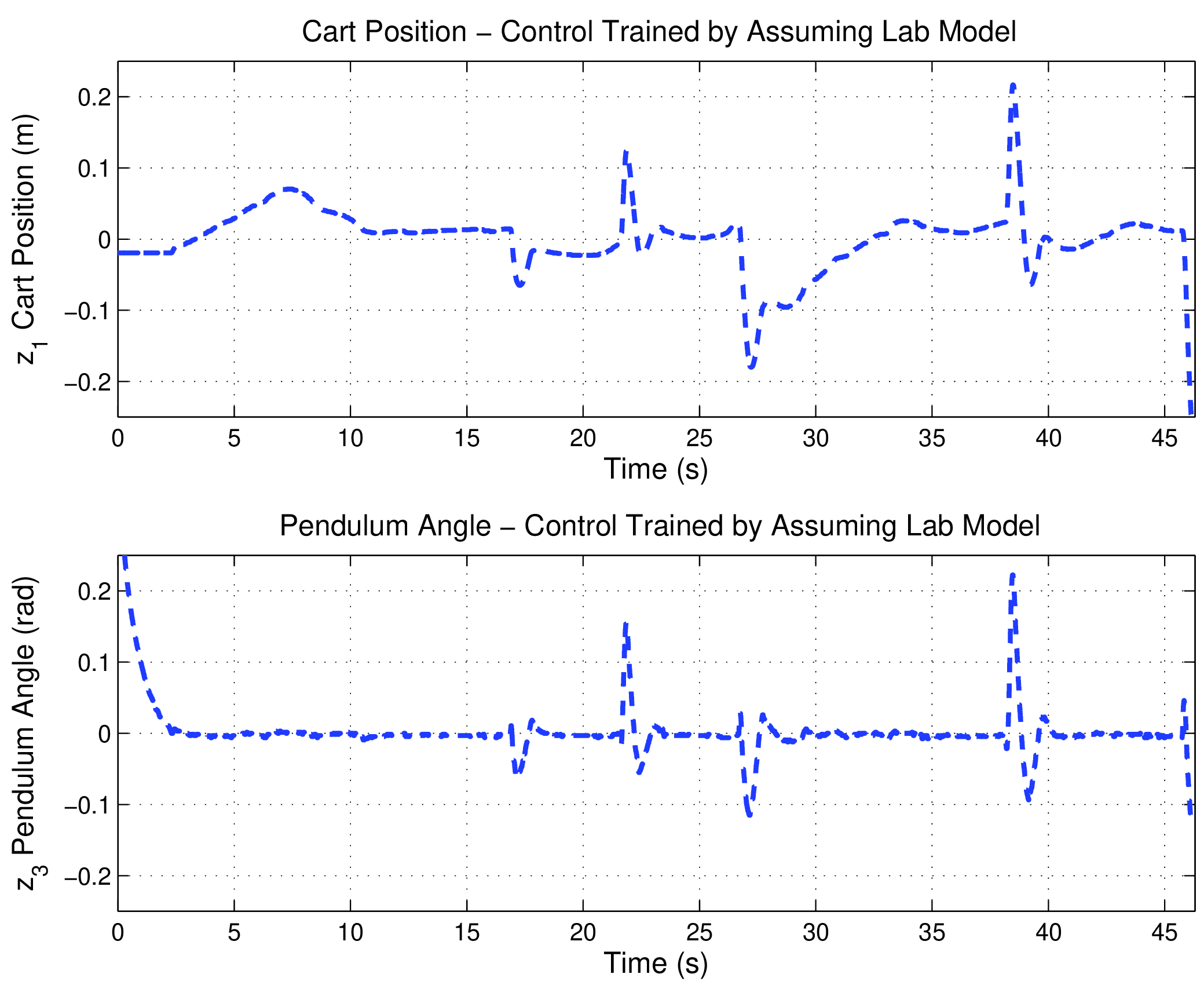}
         \caption{run \#1}
         \label{fig:NonLinControl_run1}
     \end{subfigure}
     \hspace{0.1\textwidth}
     \begin{subfigure}[b]{0.49\textwidth}
         \centering
         \includegraphics[width=\textwidth]{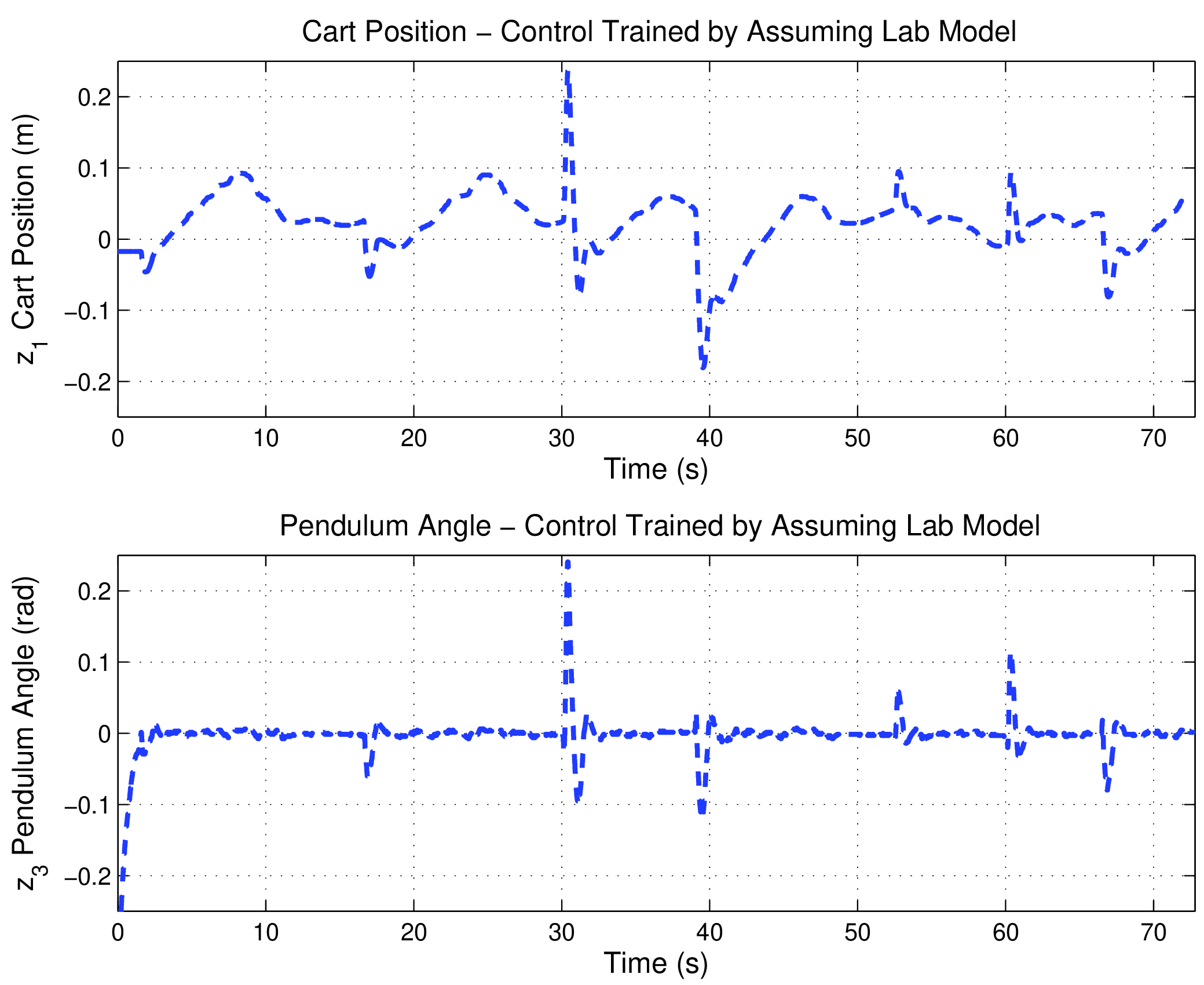}
         \caption{run \#2}
         \label{fig:NonLinControl_run2}
     \end{subfigure}
     \begin{subfigure}[b]{0.49\textwidth}
         \centering
         \includegraphics[width=\textwidth]{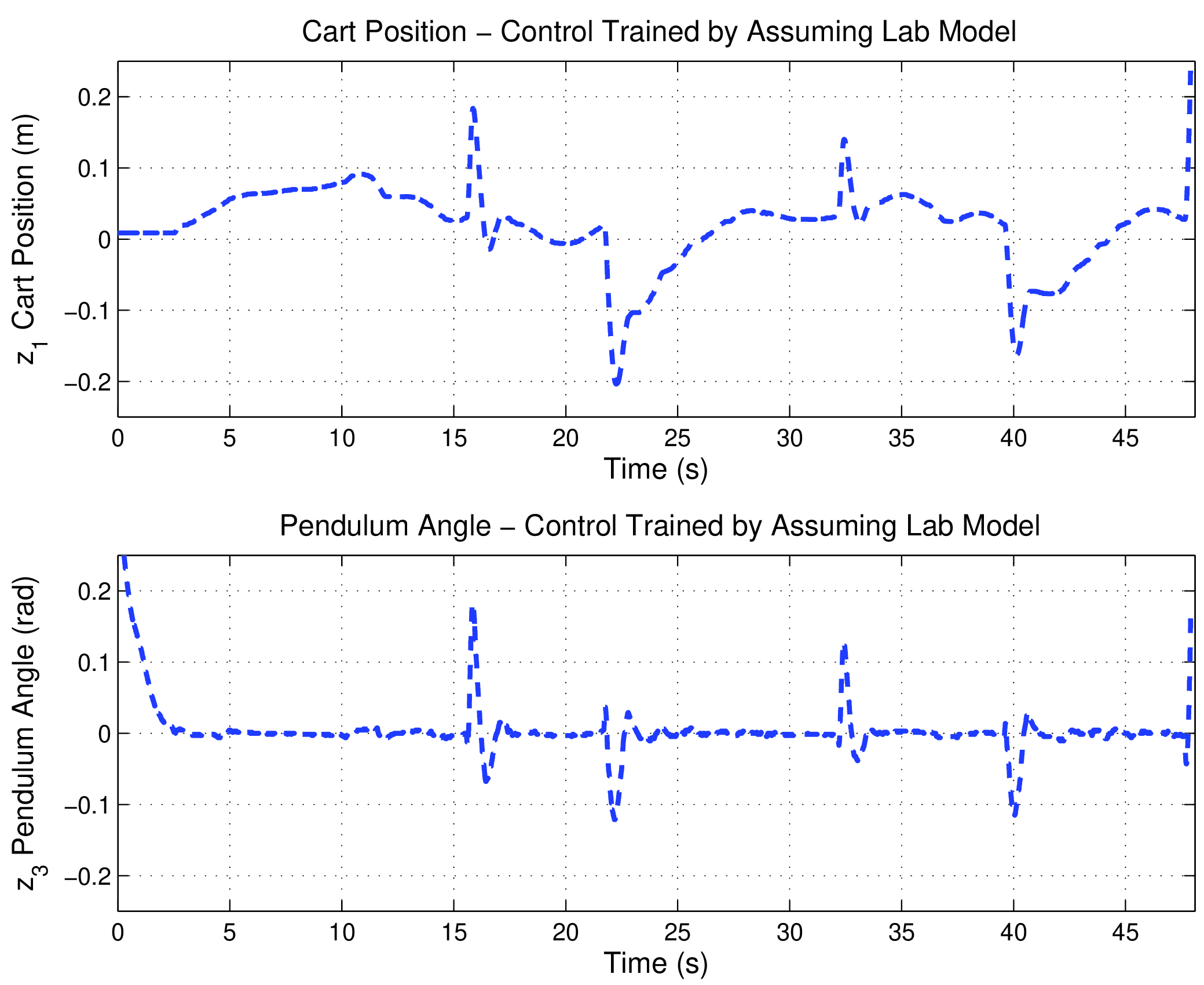}
         \caption{run \#3}
         \label{fig:NonLinControl_run3}
     \end{subfigure}
     \hspace{0.1\textwidth}
     \begin{subfigure}[b]{0.49\textwidth}
         \centering
         \includegraphics[width=\textwidth]{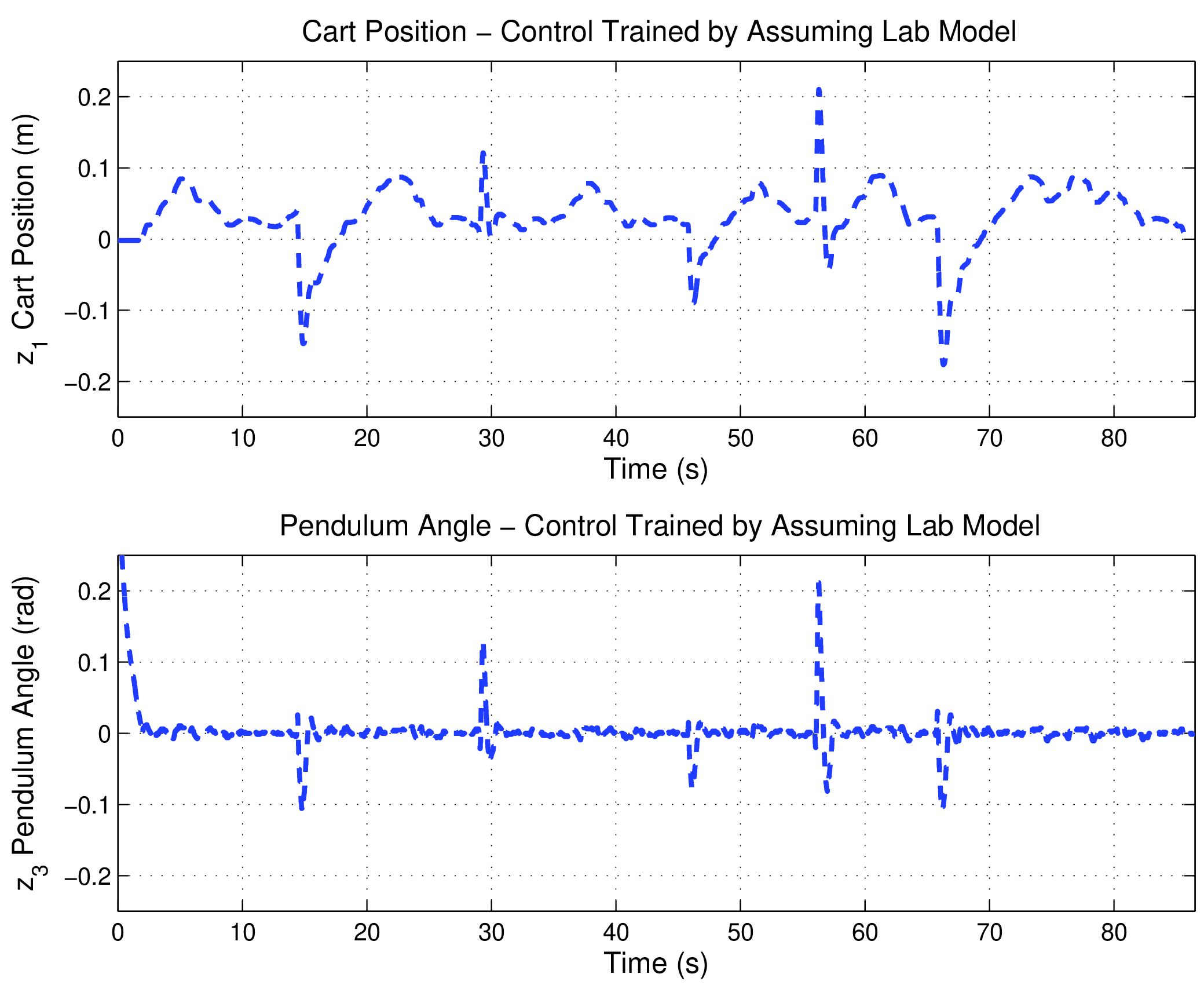}
         \caption{run \#4}
         \label{fig:NonLinControl_run4}
     \end{subfigure}
     \begin{subfigure}[b]{0.49\textwidth}
         \centering
         \includegraphics[width=\textwidth]{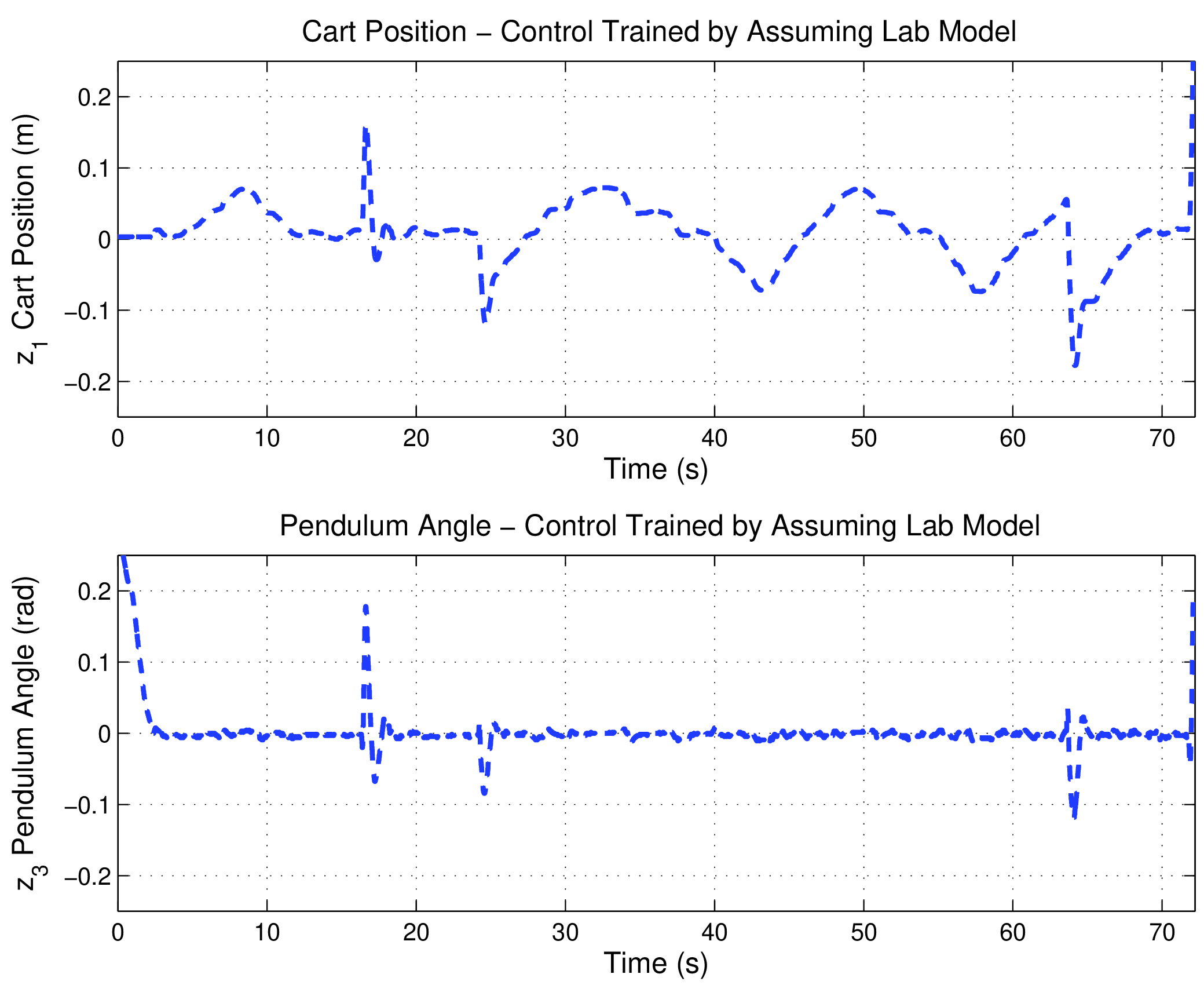}
         \caption{run \#5}
         \label{fig:NonLinControl_run5}
     \end{subfigure}
        \caption{Control learned with the lab model tested on the physical pendulum in the lab. The spikes correspond to manually tapping the pendulum to test the robustness.}
        \label{fig:FullModel_Experiment}
\end{figure}

The initial conditions that corresponded to a successful control are plotted in Figure~\ref{fig:ROA_more_samples}. The scatter plots in the off-diagonals are projections of the four-dimensional region of attraction onto the $\ds {4\choose2} = 6$ two-dimensional planes that are spanned by each pair of state variables. The diagonals are kernel density estimate (KDE) plots, which basically is a smoothed out version of histograms. It is rather clear that using the controller trained on the lab model is much more likely to actually stabilize the pendulum within the limit of the cart track, as the green region is much larger than the orange region. 
We want to point out that there are some regions close to the origin where we did not sample enough and there appears to be a ``hole'' there. This is due to our priority of pursuing the outer bounds of the region of attraction, and in our setting, the neighborhood near the origin is much easier to stabilize than the farther-away ones, so it is reasonable to assume the neighborhood near the origin are stable when they are surrounded by stable regions. 

\section{Conclusions}

Reinforcement learning is a powerful and convenient tool for modernizing controller design. However, in order to reliably deploy the control policies obtained via reinforcement learning in simulation, the training environment must align with physical hardware. From model parameters to digital control bandwidth to signal noise and delays, many elements in the real world can deteriorate the performance of a simulation-based RL control policy. We demonstrated that a stabilization control trained on a high-fidelity model could be directly deployed without any fine-tuning, and it is robust against disturbances. In contrast, the policy trained on a poor approximation was brittle in the lab despite seemingly being successful in simulation. We conducted a derivative-based sensitivity analysis to account for this discrepancy and found that a neglected damping parameter contributed significantly to the closed-loop response, and omitting it in the modeling phase resulted in failures at deployment. Moreover, the empirically estimated region of attraction estimation reveals that the mismatched system has the smallest volume, whereas the controlled system trained with a high fidelity model has a much larger basin of attraction and more robust. 

Lastly, we address some limitations of our current approach and discuss some future directions for research on deployable and reliable RL control policies. First, the study of the region of attraction revealed a potential issue of distribution shift \cite{rajeswaran_towards_2017}. Possible avenues to mitigate this issue might include distributionally robust RL \cite{iyengar_robust_2005, xu_distributionally_2012, pinto_robust_2017, nilim_robustness_nodate}, uncertainty quantification \cite{ghavamzadeh_bayesian_2015, ramos_bayessim_2019}, as well as traditional robust control theory such as $H_\infty$ control \cite{ullah_robust_2020}. Secondly, our region of attraction is empirically estimated by Monte Carlo simulation, which is very computationally expensive and memory intensive. Meanwhile, control barrier functions \cite{ames_control_2019} can be incorporated into the environment directly as a safeguard to certify the safety region for the control system \cite{pinto_robust_2017, richards_lyapunov_2018}, or trained separately as a neural network critic to help guide the policy towards certifiably safe behaviors \cite{nguyen_robust_2021, westenbroek_lyapunov_2022}. Lastly, a naive neural network design was unable to capture the inherent symmetry of the underlying dynamics. Imposing symmetry as a soft constraint or as a regularization term in the objective function may be beneficial. Although most symmetry-conscious neural network designs are for supervised learning \cite{bronstein_geometric_2021, garanger_symmetry-enforcing_2024, zhao_symmetry_2025}, RL-based control can also benefit from these structures \cite{kim_symmetric_2024}. 

\bibliography{ref_ML}

\newpage

\section{Appendix: Experimental Results}

Top/left plot of each subfigure is the cart position, and bottom/right plot of each subfigure is the pendulum angle. 

\subsection{Lab Performance of the Controller Trained with LTI Model}

\newpage
\subsection{Lab Performance of the Controller Trained with Lab Model}

\end{document}